\documentclass[fleqn]{2022AMSE}
\usepackage{bm}
\usepackage{subfigure}
\usepackage{etoolbox}

\usepackage{cite}

\begin{document}

\ensubject{Fluid Dynamics}

\ArticleType{RESEARCH PAPER}
\Year{2025}
\Vol{38}
\DOI{10.1007/s10409-022-???-?}
\ArtNo{???}
\ReceiveDate{???}
\AcceptDate{???}
\OnlineDate{???}

\title{Dynamics of an elliptical cylinder in confined Poiseuille flow under Navier slip boundary conditions}

\author[1]{Xinwei Cai}{}%
\author[1]{Xuejin Li}{}
\author[1]{Xin Bian}{bianx@zju.edu.cn}

\AuthorMark{Xinwei Cai}

\AuthorCitation{Cai, Li, and Bian}

\address[1]{State Key Laboratory of Fluid Power and Mechatronic Systems, Department of Engineering Mechanics, \\Zhejiang University, Hangzhou 310027, P. R. China}


\abstract{
A comprehensive understanding of surface wetting phenomena in microchannels is essential for optimizing particle transport and filtration processes. This study numerically investigates the dynamics of a freely suspended elliptical cylinder in confined Poiseuille flow, with a focus on Navier slip boundary conditions. The smoothed particle hydrodynamics method is employed, which is advantageous for its Lagrangian framework in handling dynamic fluid-solid interfaces with slip. Our results demonstrate that the slip conditions enable precise control over inertial focusing positions and particle motion modes. Compared to no-slip scenarios, unilateral wall slip induces two novel motion types: ``leaning" and ``rolling". When equal slip lengths are applied to both walls, even small slip values facilitate  off-center inertial focusing and elevate equilibrium positions. Slip on the cylinder surface further enhances inertial lift while suppressing rotational dynamics. In particular, under strong confinement or with large particle-surface slip lengths, we identify an additional distinct motion regime termed "inclined." These findings provide new insights for active particle manipulation in microfluidic applications.
}


\setlength{\textheight}{23.6cm}
\thispagestyle{empty}

\maketitle
\setlength{\parindent}{1em}

\vspace{-1mm}
\begin{multicols}{2}

\section{\label{sec:level1}Introduction}
\noindent Flows involving solid particles in suspension are ubiquitous across industrial and biomedical applications, spanning from transport and filtration in micro- and nanofluidics~\cite{Kim1991, Xue2025} to targeted drug delivery in blood flow~\cite{mitchell2021engineering,qi2021quantitative}. While dynamics of spherical particles have been extensively studied~\cite{Kim1991, segre1961, martel2014inertial, Vazquez-Quesada2015c, alexeev2024}, behaviors of non-spherical particles in flows are less understood. Based on pioneering works~\cite{seki1993, feng1995unsteady}, we previously conducted a numerical study on the dynamics of an elliptical cylinder in a confined Poiseuille flow between two parallel plates under no-slip boundary conditions~\cite{cai2024_PartA_noslip}. 
We identified three distinct types of periodic motion in the steady state at small Reynolds numbers: tumbling, oscillation with the major axis aligned with the flow direction, and oscillation with the minor axis aligned with the flow direction. In weakly confined channels, the particle consistently tumbles and exhibits fixed off-center focusing positions, driven primarily by the competition between shear gradient lift and the wall-induced force in the transverse direction~\cite{martel2014inertial}. In contrast, in strongly confined channels, the particle undergoes continuous oscillations along the centerline, with its behavior determined by Reynolds number, initial conditions, and aspect ratios of the particle~\cite{cai2024_PartA_noslip}. In this work, we extend our numerical investigation to include slip boundary conditions, encompassing both wall slip and particle surface slip, which are prevalent in a wide range of micro- and nanoscale flows. 

The no-slip hypothesis, a cornerstone of fluid mechanics for over two centuries, asserts that the fluid layer adjacent to a solid surface moves with zero relative velocity with respect to the surface. This principle has been widely applied in macroscale continuum flows~\cite{Neto2005_et_al_RepProgPhys}. However, the no-slip condition is not universally valid, particularly at the micro- and nanoscale, where slip at fluid-solid interfaces is often observed. For instant, flows in hydrophobic capillaries~\cite{Rothstein2010_AnnRevFluidMech} and through a membrane composed of aligned carbon nanotubes~\cite{Majumder2005_et_al_Nature, holt2006fast} demonstrate significant interfacial slip. The concept of slip boundary conditions was first proposed by Navier~\cite{navier1823memoire} and has since been extensively investigated~\cite{maxwell1879vii,ou2004laminar,sanders2006bubble}. In micro- and nanofluidic systems, accounting for slip velocity at interfaces is often essential for accurately modeling flow behavior.

Over the years, slip boundaries have been the focus of extensive research and practical applications.
They can be categorized into several key areas. First, drag reduction is one of the most significant and widely studied applications. For example, Papadopoulos et al.~\cite{papadopoulou2021nanopumps} used molecular dynamics (MD) simulations to demonstrate that carbon nanotubes with wettable surface patterns can act as nanopumps, enabling ultrafast water transport without external pressure gradients. Qu\'er\'e~\cite{quere2005non} reported the developement of superhydrophobic materials that impart non-adhesive properties to droplets. Ou et al.~\cite{ou2004laminar} experimentally showed that superhydrophobic surfaces can reduce flow resistance by up to $40\%$ in laminar flow. Additionally, Secchi et al.~\cite{secchi2016massive} found that flow rates through carbon-based nanoporous membranes can exceed classical predictions by up to an order of magnitude. Second, slip boundaries enable the modification of flow field distributions, facilitating active flow control. Xuan et al.~\cite{xuan2022active} used lattice Boltzmann method to manipulate the inertial focusing positions of particles by ajusting a unilateral wall slip length in a Poiseuille flow. 
Muralidhar et al.~\cite{muralidhar2011influence} experimetally investigated the effect of superhydrophobic-induced slip on flow past a cylinder, observing delayed onset of K\'arm\'an vortex shedding and an extended recirculation region in the wake. Mastrokalos et al.~\cite{mastrokalos2015optimal} proposed an optimization framework, guided by finite volume method simulations,  to partially or completely suppress vortex streets with minimal control effort. Finally, slip boundaries play a crucial role in enhancing nanofiltration processes. For instance, liquid slip has been utilized to improve electroosmotic flow in nanomedia, leading to higher energy conversion efficiency~\cite{sparreboom2009principles,rezaei2018surface}.

Furthermore, slip is known to reduce the hydrodynamic stress exerted by shearing liquids on solid particle surfaces, leading to decreased translational and rotational drag coefficients for both spherical and non-spherical particles~\cite{luo2008effect, sellier2013arbitrary}. Specifically, slip retards rotational dynamics of freely suspended spherical particles in simple shear flow~\cite{youngren1975rotational, keh2008slow, sherwood2012resistance}. Kamal et al.~\cite{kamal2020hydrodynamic}, using a combination of MD method and continuum simulations, demonstrated that a thin nanoplatelet aligns at a small angle relative to the flow direction when a large slip length is specified on its surface,  deviating from the full tumbling motion predicted by classical Jeffery orbits~\cite{jeffery1922}.

It is apparent from the preceding discussions that there lacks an investigation on the dynamics of non-spherical particles in confined channel flows, with possible slip boundary conditions at the solid-fluid interfaces. To fill in this research gap, we employ smoothed particle hydrodynamics~(SPH) method
for its Lagrangian nature to steadily handle dynamics interfaces. Originally developed for astrophysics, SPH has since been extensively adapted and widely applied to various fluid dynamics problems~\cite{Monaghan2012,le2025smoothed, zhang2017},
especially for hydrodynamic interactions between multiple solid particles and confining walls~\cite{Bian2012,Bian2014,Vazquez-Quesada2015c, cai2024simulating}. Based on the treatment of arbitrary slip length for fluid-solid interfaces
of arbitrary geometry~\cite{cai2023_li_xin_JCompPhys}, we can accurately capture the partial-slip boundaries between the elliptical particle and the fluid, as well as between the solid walls and the fluid. 

The structure of the rest is as follows. In Section~\ref{method} we describe the numerical method. In Section~\ref{results} we present our results and discussions. Finally, we summarize this work in Section~\ref{conclusion}.

\section{The Method}\label{method}

\subsection{Lagrangian hydrodynamic equations, rigid body motion and boundary conditions}
We consider a Newtonian fluid governed by the continuity and Navier-Stokes equations in Lagrangian form as follows
\begin{eqnarray}
   \frac{{\mathrm{d}}\rho}{{\mathrm{d}} t} & =&  -\rho \nabla \cdot \mathbf{v},\label{continuity_eq} \\
    \rho \frac{{\mathrm{d}}\mathbf{v}}{{\mathrm{d}}t} &=& -\nabla p + \eta \nabla^2\mathbf{v} + \rho\mathbf{f}, \label{ns_eq}
\end{eqnarray}
where $\rho$, $\mathbf{v}$, $p$, $\eta$, and $\mathbf{f}$ are material density, velocity, pressure, dynamic viscosity and body force per unit mass, respectively.
To colse the weakly compressible description, an equation of state relating the pressure to the density is required
\begin{eqnarray}
p=c_s^2(\rho-\rho_0)+\chi
\label{eoc_eq},
\end{eqnarray}
where $\rho_0$ is the equilibrium density. $c_s$ is an artifical sound speed chosen based on a scale analysis such that the pressure field reacts strongly to small deviations in the density, thus satisfying quasi-incompressibility. Here, $\chi = c_s^2\rho_0$ is a positive constant introduced to enforce the non-negativity of the pressure on discrete particles. 

The movement and rotation of rigid body are described by Newton and Euler equations
\begin{eqnarray}
\mathbf{F}(t) &=& M\frac{d\mathbf{U}(t)}{dt},\\
\mathbf{T}(t) &=& \mathbf{I}\cdot\frac{d \bm{\Omega} (t)}{dt} + \bm{\Omega}(t)\times[\mathbf{I}\cdot \bm{\Omega}(t)],
\end{eqnarray}
where $\mathbf{I}$, $\bm{\Omega}$ and $\mathbf{T}$ represent the inertial tensor, the angular velocity and the
torque exerted on the moving solid in the body-fixed coordinate system, respectively. 

The Navier slip boundary condition applies in this work, which is given as follows
\begin{eqnarray}
v_{s}= L_s \frac{\partial v^{\tau}}{\partial n}\label{v_slip}.
\end{eqnarray}
Here $v_s$ is the slip velocity at the surface, $L_s$ is the slip length characterizing the slip boundary, $\frac{\partial v^{\tau}}{\partial n}$ is gradient of velocity in the tangential direction at the interface, respectively. The slip length $L_s$ is equivalent to the ratio of the dynamic viscosity and surface friction coefficient, where $L_s = 0$ and $L_s \to \infty$ are two special cases corresponding to the no-slip and free-slip, respectively~\cite{cai2023_li_xin_JCompPhys}.


\subsection{The SPH method}

The continuity equation is discretized as follows~\cite{Monaghan2005}
\begin{eqnarray}
\frac{d \rho_{i}}{d t} &=& \rho_{i}\sum_j \frac{m}{\rho_j}\mathbf{v}_{ij} \cdot \frac{\partial{W}}{\partial{r_{ij}}}\mathbf{e}_{ij},\label{drho}
\end{eqnarray}
where $\rho_j$ is the density of particle $j$ and $m$ is the identical mass for each particle. The relative position and velocity of particles $i$ and $j$ are represented by $\mathbf{r}_{ij}$ and $\mathbf{v}_{ij}$, respectively. 
Additionally, $r_{ij}=|\mathbf{r}_{ij}|$ and $\mathbf{e}_{ij}=\mathbf{r}_{ij}/r_{ij}$. 
We employ the quintic spline as the kernel function,
which vanishes beyond a cutoff radius $r_c$.
After several time steps, numerical errors in the density integration can accumulate; hence, we restore the density every $10$ time steps via the summation equation
\begin{eqnarray}
    \rho_i = \sum_j m_j W_{ij}.
\end{eqnarray}
We note that the results obtained in this manner are consistent with those obtained by restoring the density at each step, but the former significantly reduces computational effort.

To minimize numerical errors due to irregular distributions of SPH particles, we adopt the transport velocity method. The discrete momentum equation is given by~\cite{cai2023_li_xin_JCompPhys, Adami2013}
\begin{eqnarray}
\frac{\tilde{d}\mathbf{v}_i}{d t} &=& \frac{1}{m_i}\sum_j(\frac{1}{\sigma_i^2}+\frac{1}{\sigma_j^2})[\mathbf{f}_p + \mathbf{f}_\mu + \mathbf{f}_A ] + \mathbf{g}_i\label{mom_equ}.
\end{eqnarray}
Here $\tilde{d}(*)/dt$ denotes a new definition of the material derivative. $\sigma$ is the number density, defined as the ratio of $\rho$ to mass $m$. $\mathbf{f}_p$, $\mathbf{f}_\mu$ and $\mathbf{f}_A$ represent the pressure term, the viscosity term, and the additional term due to convection, respectively:
\begin{eqnarray}
\mathbf{f}_p &=& -\frac{\rho_jp_i+\rho_ip_j}{\rho_i+\rho_j}\frac{\partial{W}}{\partial{r_{ij}}}\mathbf{e}_{ij},\\
\mathbf{f}_\mu &=& \frac{2\mu_i\mu_j}{\mu_i+\mu_j}\frac{\mathbf{v}_{ij}}{r_{ij}}\frac{\partial{W}}{\partial{r_{ij}}},\label{viscosity_eq}\\
\mathbf{f}_A &=& \frac{1}{2}(\mathbf{A}_i+\mathbf{A}_j)\cdot\frac{\partial{W}}{\partial{r_{ij}}}\mathbf{e}_{ij}.
\end{eqnarray}
Here, we use the density-weighted pressure and the interparticle-averaged shear viscosity. $\mathbf{A}=\rho\mathbf{v}(\mathbf{\tilde{v}-\mathbf{v}})$ is a tensor from the dyadic product of the two vectors. Additionally, $\mathbf{v}$ is the velocity for momentum and force calculations, while $\mathbf{\tilde{v}}$ is the modified transport velocity for updating the position of each fluid particle. The discrete form of $\tilde {\mathbf{v}}$ is computed as
\begin{equation}
\tilde{\mathbf{v}}_i = \mathbf{v}_i(t) + \delta t(\frac{\tilde{d}\mathbf{v}_i}{dt}-) \frac{\chi}{m_i}\sum_j(\frac{1}{\sigma_i^2}+\frac{1}{\sigma_j^2})\frac{\partial{W}}{\partial{r_{ij}}}\mathbf{e}_{ij}),
\end{equation}
where $\delta t$ is the time step, and the positive constant $\chi$ only appears here, not in the momentum equation.

\subsection{Boundary treatment at fluid-solid interface}
\begin{figure*}
\centering
\includegraphics[scale=1]{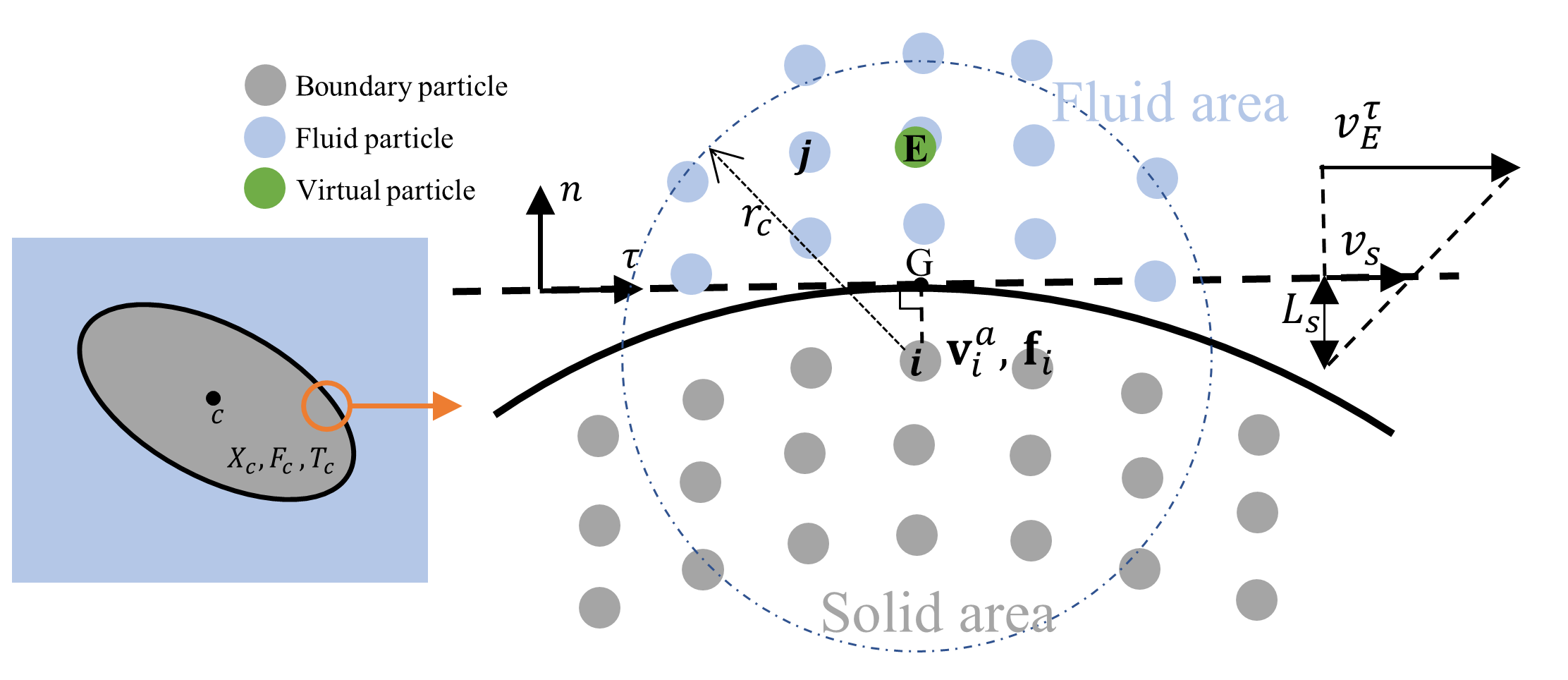}
\caption{Schematic of the boundary treatment. Given a boundary particle $i$, an interface plane tangential to the solid surface is defined to be
perpendicular to its normal direction $\mathbf{n}$, and intersects at $G$.
A virtual particle $E$ represents the average effect of each neighboring fluid particle $j$ on the boundary particle $i$.
An artificial velocity $\mathbf{v}^{a}_i$ on boundary particle $i$ is calculated based on the velocity ${\bf v}_E=(v^{\tau}_E, v^n_E)$ of particle $E$ and the slip length $L_s$ at the interface, while taking into account distances of particles $E$ and $i$ to the interface.
The interface has a slip velocity $v_s$ at $G$. The total force of the neighboring fluid particles on the boundary particle $i$ is ${\bf f}_i$.}
\label{boundary_sph}
\end{figure*}

Boundary treatment is always a challenge in SPH method. Here we use a special boundary method to deal with the no-/partial slip at the fluid-solid interface~\cite{cai2023_li_xin_JCompPhys}. Boundary particles are used to represent the wall and ellipse, which have the same mass and resolution $\Delta x$ as the fluid particles, as depicted in Fig.~\ref{boundary_sph}.

To implement the slip boundary condition at the interface between a rigid body and the fluid, we assign to each boundary particle $i$ an artificial velocity, which is determined by the properties of a virtual particle $E$ and a prescribed slip length $L_s$ at the interface~\cite{cai2023_li_xin_JCompPhys}.
The slip length $L_s$ at the interface is an input parameter:  $L_s = 0$ represents the no-slip condition, while $L_s > 0$ indicates a partial-slip condition. 
The position $\mathbf{x}_E$ and velocity $\mathbf{v}_E$ of the virtual particle $E$ are calculated using SPH interpolations, which represents an average effect due to neighboring fluid particles on the boundary particle $i$, as shown in Fig.~\ref{boundary_sph}. 
Thereafter, an artificial velocity $\mathbf{v}_i = (v_i^\tau, v_i^n)$ is assigned to the boundary particle $i$ as
\begin{eqnarray}
v_i^\tau & = & \frac{L_s - d_i}{L_s + d_E}v_E^\tau + \frac{d_E + d_i}{L_s + d_E}v_G^\tau, \\
v_i^n &=& -\frac{d_i}{d_E}(v_E^n - v_G^n) + v_G^n. 
\end{eqnarray}

Here $\tau$ and $n$ represent the tangential and normal directions at the interface plane, respectively. $d_i$ and $d_E$ correspond to the distances of boundary particle $i$ and virtual particle $E$ from the interface, respectively. The velocity of the intersection point at the interface, denoted as $\mathbf{v}_G = (v_G^\tau,v_G^n)$, can be determined from the motion of a rigid body.
It is important to note that the artificial velocity of particle $i$ is solely utilized in the calculation of the viscous force in Eq.~(\ref{mom_equ}), and it is not intended for the kinematics of the elliptical cylinder. To calculate the density $\rho_i$ of particle $i$, a similar approach is employed to satisfy the pressure boundary condition at the interface. For more detailed information, we recommend referring to previous works~\cite{cai2023_li_xin_JCompPhys, Adami2012}.

With appropriately implemented boundary conditions, it becomes possible to calculate the total force $\mathbf{F}_c$ and torque $\mathbf{T}_c$ acting on a moving rigid body. These quantities are obtained by summing the forces of all the SPH boundary particles that make up the rigid body. The linear velocity $\mathbf{V}_c$, the angular velocity $\bm{\Omega}_c$ and the position $\mathbf{X}_c$ of the rigid body can then be updated accordingly. Furthermore, the positions of the SPH boundary particles inside the rigid body can be updated based on the motion of the rigid body. For a fuller understanding of the kinematics involved, please refer to the solid particle dynamics described in reference~\cite{cai2024_PartA_noslip}.

\subsection{Particle–wall interaction}
As an solid particle approaches a wall, the lubrication effects of the fluid within the thin gap between the particle and the wall precludes unphysical overlap. However, accurately capturing the lubrication hydrodynamics necessitates a very fine SPH resolution, rendering it computationally expensive. 
The situation deteriorates in the presence of a unilateral slip wall,
where the no-slip wall "pushes" the particle even closer to the side of slip wall.
With usual resolutions, nonphysical overlaps emerge as there are insufficient SPH particles between the particle surface and the wall. Therefore, two remedies are called for: one is to switch on SPH interactions between solid boundary particles and wall boundary particles as they were fluid particles; the other is to introduce a short-range repulsive force\cite{xia2009_chen}. 

The repulsive force that arises during particle-wall collision is defined as follows:
\begin{eqnarray}
        \mathbf{f}_r = \frac{C_m}{\epsilon}(\frac{|\mathbf{x}_e - \mathbf{x}|-dr}{dr})^2 \frac{\mathbf{x}_e-\mathbf{x}}{|\mathbf{x}_e-\mathbf{x}|}
\end{eqnarray}
where $\mathbf{x}_e$ is the point on the elliptical particle closest to the wall and $\mathbf{x}$ is the corresponding point on the wall. $C_m=\frac{MU_{max}^2}{2a}$ is the the characteristic force, $\epsilon = 10^{-4}$, $dr$ is the force range set to be $\Delta x$. Note that this force may produce torque for an elliptical particle as it is not necessarily directed through the center of mass.

The SPH interaction is activated when a SPH boundary particle on the elliptical is less than a cutoff radius $r_c$ from the wall. Additionally, the short-range repulsive force is activated when this distance is less than an initial particle distance $\Delta x$. In many scenarios, the first interaction alone is not sufficient, so the short-range repulsive force is also activated.

\section{Results and discussions}\label{results}
\begin{figure*}[!htb]
\centering
\includegraphics[scale=0.6]{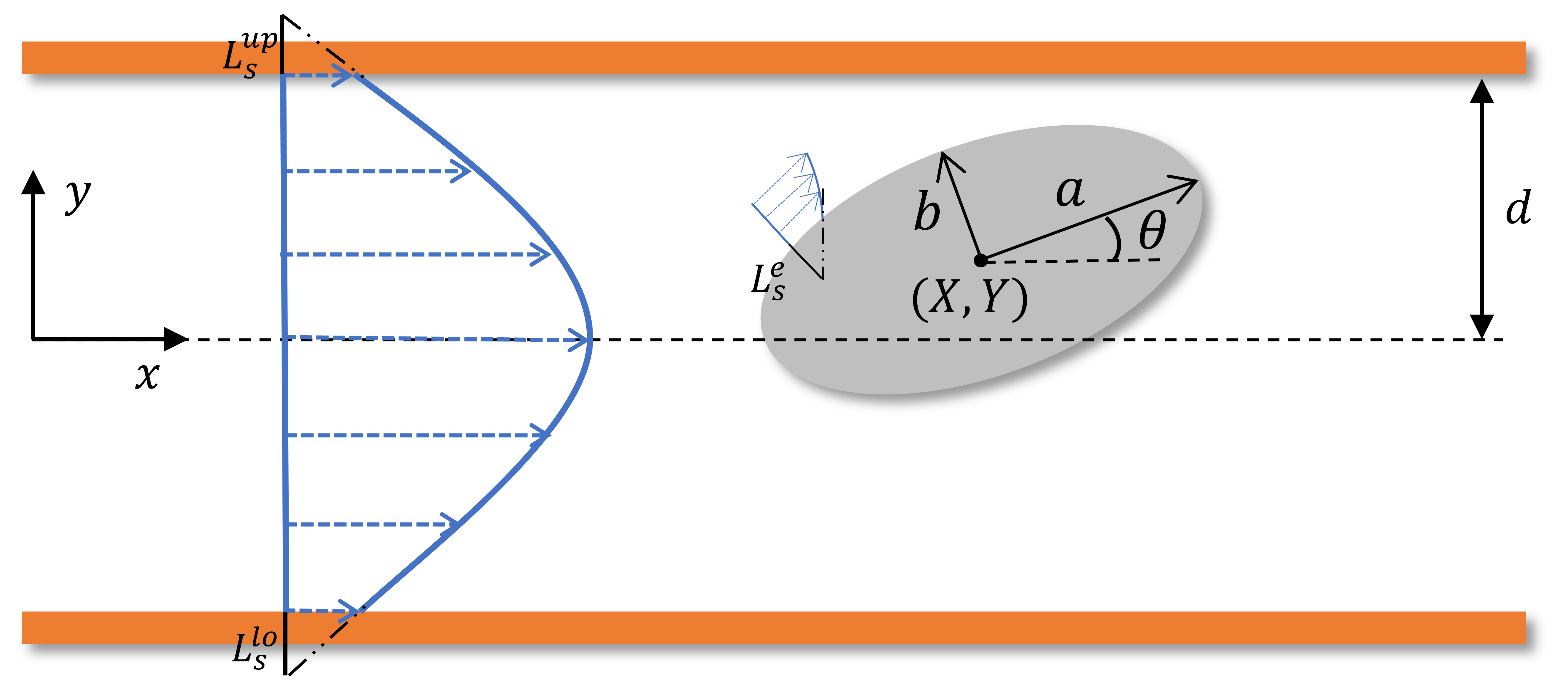}
\caption {The geometry of an elliptical cylinder in Poiseuille flow. $a$ and $b$ are the major and minor semi-axes of the ellipse. $\theta$ is the angle between the major axis and the $x$-direction. $d$ is the half-width of the channel.  Solid-fliud interfaces have slip boundary conditions, with $L_s^{up}$, $L_s^{lo}$ and $L_s^e$ representing the slip lengths on surfaces of the upper wall, the lower wall and the elliptical cylinder, respectively.}
\label{sketch_ellipse}
\end{figure*}

The configuration of the simulations is shown in Fig.~\ref{sketch_ellipse},
where a neutrally buoyant elliptical cylinder with semi-axes $a$ and $b$ is confined by two parallel walls.
A constant body force ${\bf f}_b$ is applied along the $x$ direction to drive
a Poiseuille flow, which has periodic boundaries in the flow direction.
Solid-fluid interfaces have slip boundary conditions,
with $L_s^{up}$, $L_s^{lo}$ and $L_s^e$ representing the slip lengths on the surfaces of the upper wall, the lower wall and the particle, respectively.
The centerline of the channel is at $y=0$ and the half width of the channel is denoted by $d$. The aspect ratio of the ellipse is $\alpha=a:b=2$, while the channel-particle size ratio is defined by $\beta=d:a$. The characteristic length is set to $2a = 1$. The density $\rho$ and the kinematic viscosity $\nu$ are set to $1$ so that the Reynolds number is defined as 
\begin{eqnarray}
    Re = \frac{2U_{0}a}{\nu} = \frac{f_bd^2a}{\nu^2}.
\end{eqnarray}
Here, $U_0$ is the maximum velocity magnitude of the fluid in the absence of the ellipse with no slip boundary condition on both walls and $f_b$ is the corresponding body force.
When slip boundaries are applied,
the body force remains constant for a given Reynolds number
and adjusts to achieve different Reynolds numbers. 
The position of the center of the ellipse and the orientation are denoted by $(X,Y)$ and $\theta$. We set the channel sufficiently long with $L = 10a$ to eliminate any effects due to the periodic boundaries. 
We adopt a resolution of $20$ SPH particles along the minor semi-axis of the ellipse, that is, $\Delta x = b/20$, which has been shown to be sufficient~\cite{cai2024_PartA_noslip}. 


\begin{figure*}
	\centering  
	\subfigure[Different slip lengths $L_s$ specified on the upper wall and no-slip on the lower wall.]{
		\label{upper_slip_poiseuille_vel}
		\includegraphics[scale=0.5]{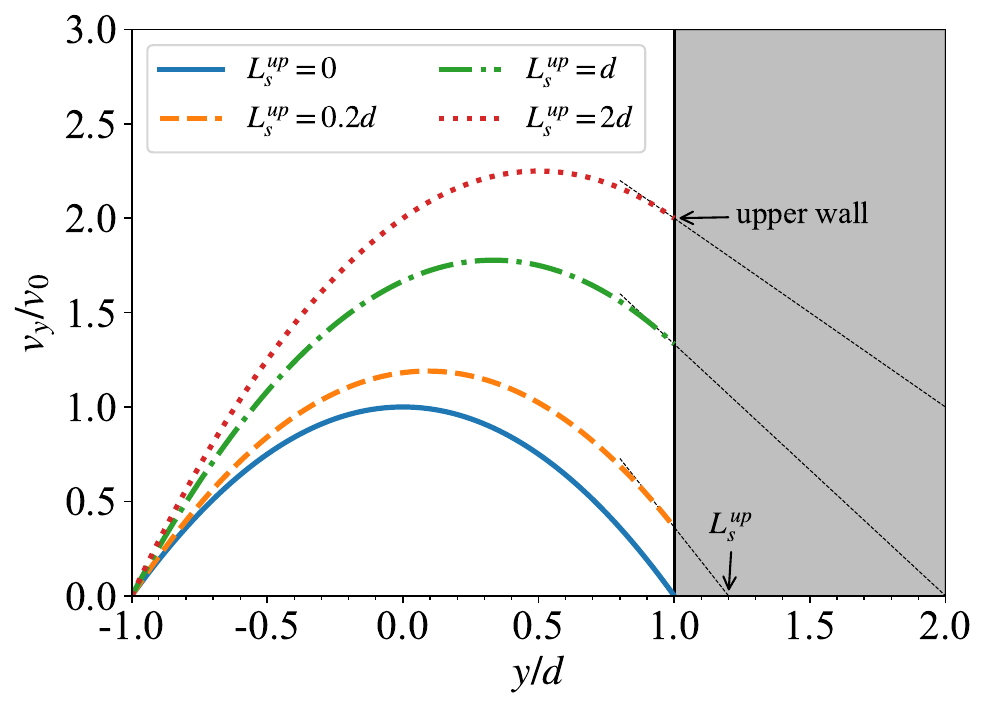}}
	\subfigure[Identical slip lengths $L_s$ are specified on the both walls.]{
		\label{both_wall_slip_poiseuille_vel}
		\includegraphics[scale=0.5]{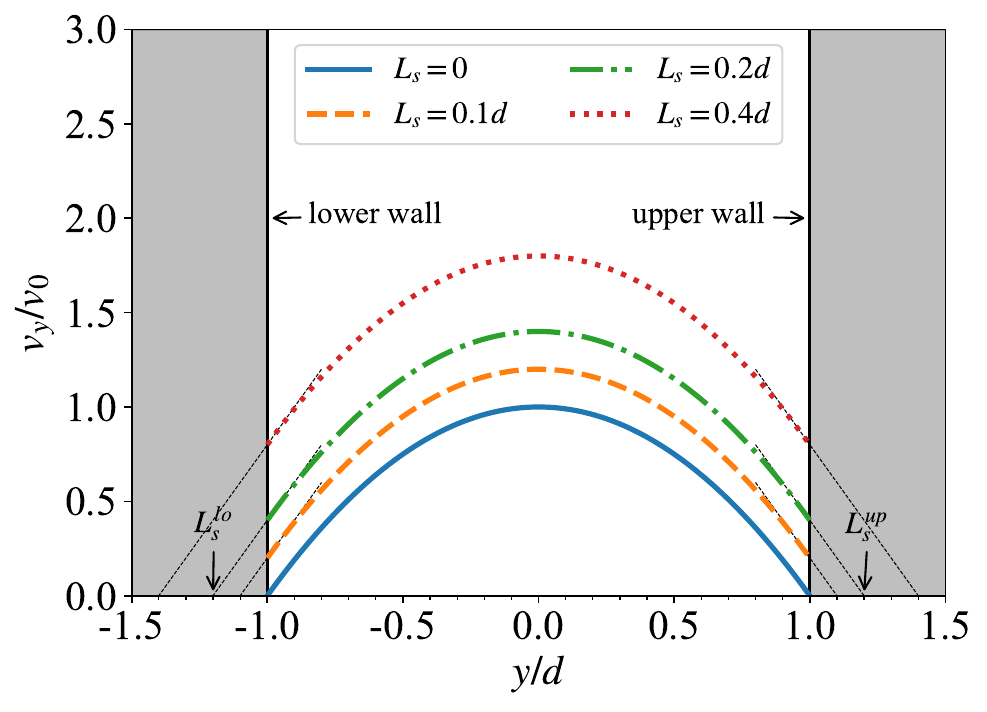}}
	\caption{Velocity profiles of the Poiseuille flow with 
different slip lengths on the wall. The dashed line indicates the linear extrapolation of the velocity at the boundary, and the distance from the intersection with the horizontal axis to the boundary represents the slip length.}
	\label{slip_poiseuille_vel}
\end{figure*}

In the absence of the elliptical particle,  when slip lengths are specified on the walls the velocity distribution of the fluid at steady state can be expressed as follows:
\begin{eqnarray}
     v_x(y)&=&-\frac{f_b}{2\nu}(y+d)^2+\frac{2f_bd(L_s^{up}+d)(y+d)}{\nu(2d+L_s^{lo}+L_s^{up})}\notag \\
     &+&\frac{2f_bL_s^{lo}d(L_s^{up}+d)}{\nu(2d+L_s^{lo}+L_s^{up})},\label{poiseuille_state_vel}
\end{eqnarray}
where $L_s^{up}$ and $L_s^{lo}$ represent the slip lengths on the upper and lower walls, respectively.
Introducing slip lengths at the walls modifies the velocity profile and potentially breaks the symmetry of the flow. 
Specifically, we consider two scenarios. The first is where the upper wall has slip boundary condition and the lower wall remains no-slip. How a few representative slip lengths on the upper wall alter the velocity distribution is illustrated in Fig.~\ref{upper_slip_poiseuille_vel}. In the second scenario, 
different identical slip lengths specified on both walls 
affect the velocity profiles as shown in Fig.~\ref{both_wall_slip_poiseuille_vel},
where flow profiles remain symmetric.

In the third scenario,  both walls have no-slip boundary condition,
but the surface of the elliptical particle has different slip lengths,
which affects its lag velocity and surface stress. 

For the results of each combination of parameters~($\beta$, $L^{up}_s$, $L^{lo}_s$, $L^{e}_s$, $Re$), we always start each simulation with different initial configurations of the elliptical cylinder, namely, different
$(X_0, Y_0)$ and $\theta_0$ so that multiple steady states can be effectively explored.



\subsection{Slip length on the upper wall}

We start with the first scenario where only the upper wall is assigned a slip condition. We select six values of channel-particle size ratio $\beta$:  $1.1$, $1.2$, $1.4$, $1.6$, $1.8$, $2.0$, twelve values of the slip length on the upper wall $L_s^{up}/d$: $0.1$, $0.2$, $0.3$, $0.4$, $0.6$, $0.8$, $1.0$, $1.2$, $1.4$, $1.6$, $1.8$, $2.0$ and two values of Reynolds number $Re$: $10$ and $1$. 

\begin{figure*}
	\centering  
	\subfigure[$Re=10$]{
		\label{upper_slip_phaseRe_10}
		\includegraphics[scale=0.45]{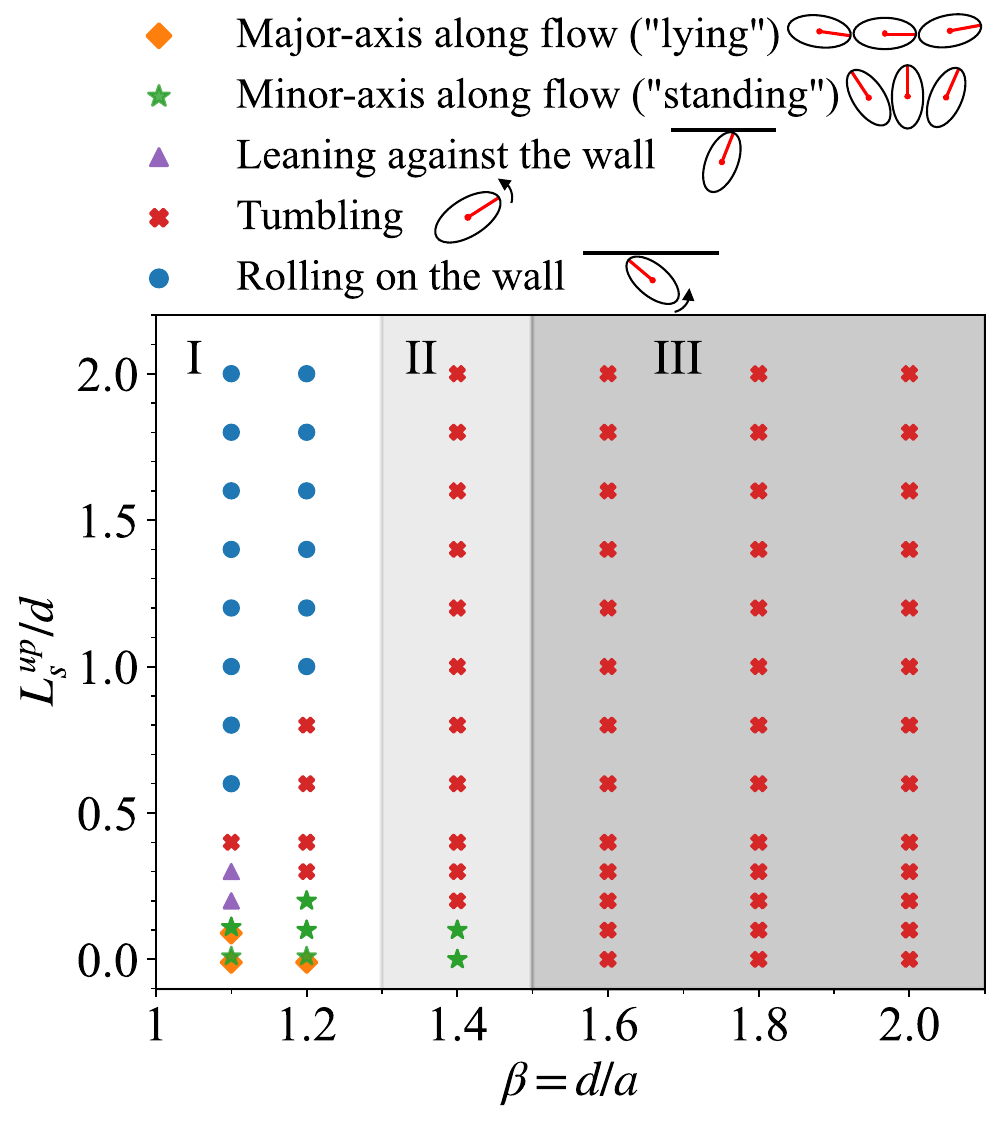}}
	\subfigure[$Re=1$]{
		\label{upper_slip_phaseRe_1}
		\includegraphics[scale=0.45]{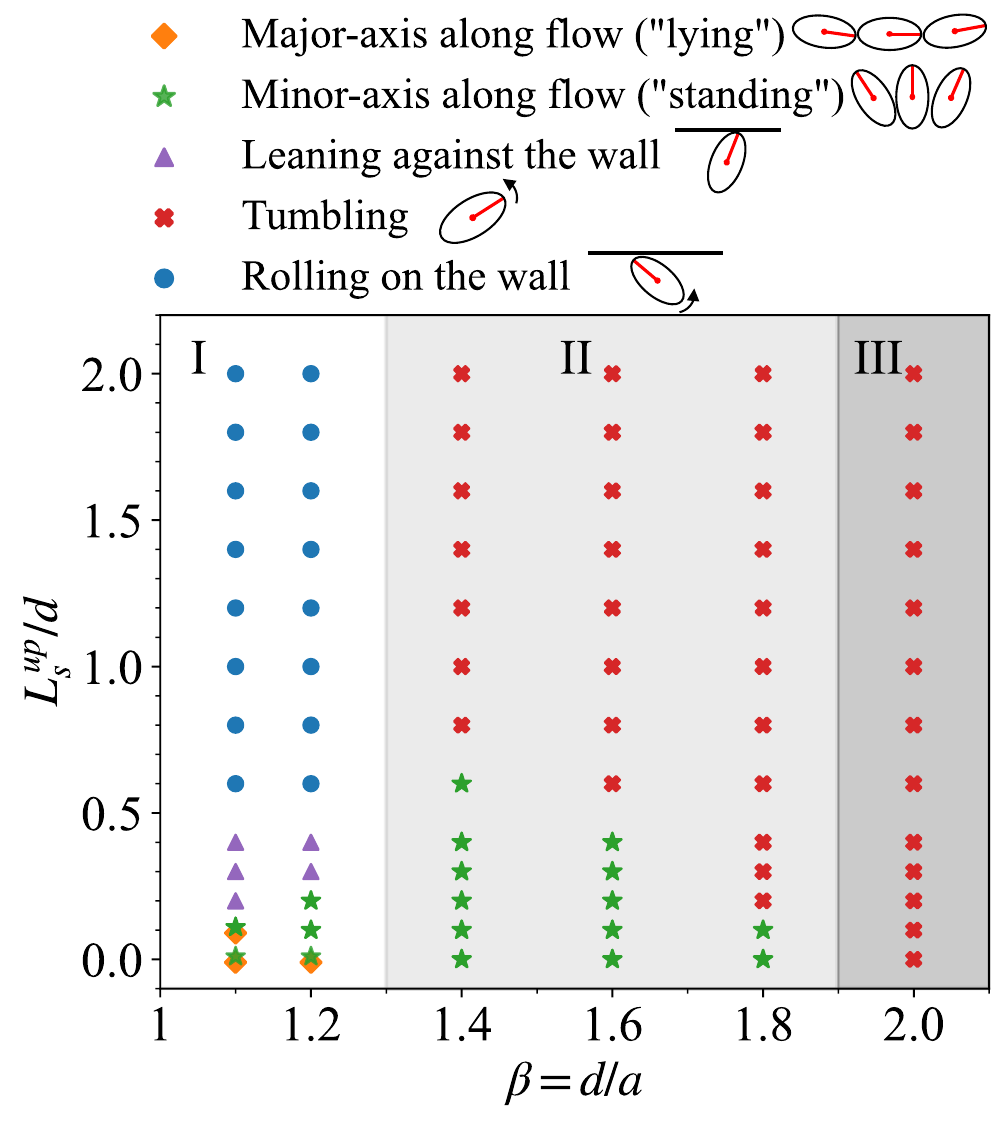}}
	\caption{Phase diagram for the steady motion types of the ellipse with
respect to the channel–particle size ratio $\beta$ and slip length on the upper wall $L_s^{up}/d$ for two Reynolds numbers. }
	\label{upper_slip_phase}
\end{figure*}

\begin{figure*}
	\centering  
	\subfigure[$Re=10$]{
		\label{}
		\includegraphics[scale=0.45]{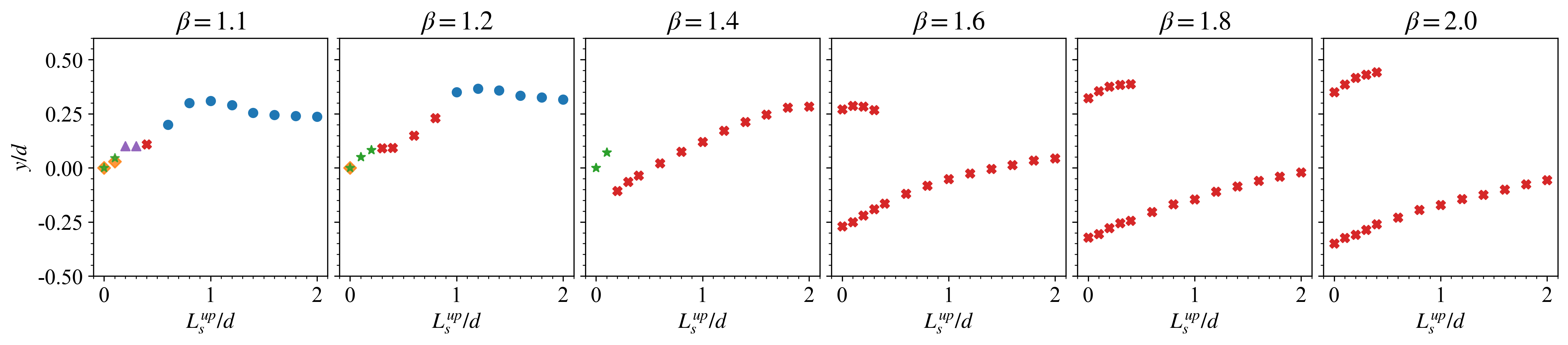}}
	\subfigure[$Re=1$]{
		\label{}
		\includegraphics[scale=0.45]{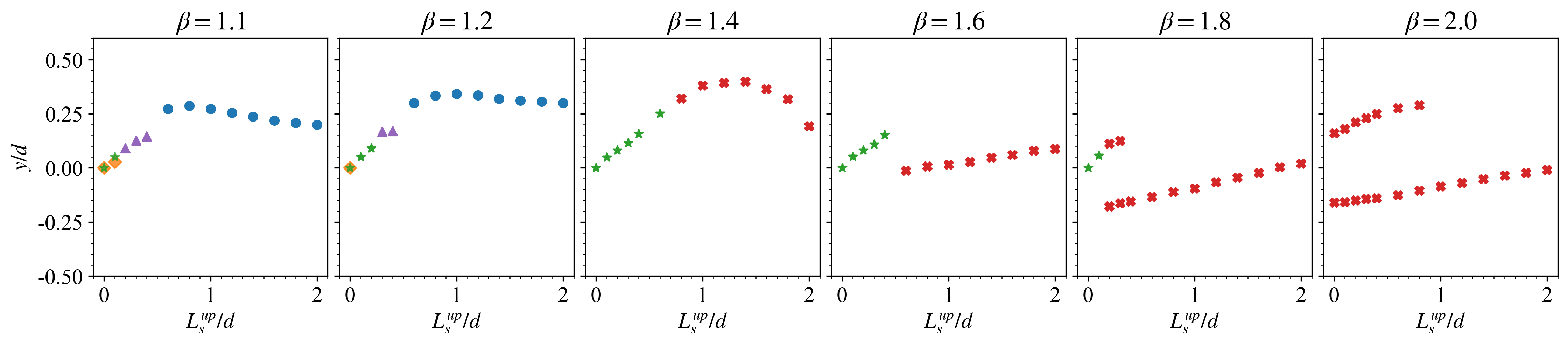}}
	\caption{Equilibrium positions of the center of mass of the ellipse with respect to the channel-particle size ration $\beta$ and slip length on the upper wall $L_s^{up}/d$ for two Reynolds numbers. The scatter symbols indicate the types of motion and have the same denotation as illustrated in Fig~\ref{upper_slip_phase}. Since the equilibrium position exhibits periodic oscillations in the transverse direction, the average value is taken for clarity.}
	\label{upper_slip_equilibrium_position}
\end{figure*}

Applying slip length to the unilateral wall in Poiseuille flow alters both the type of motion and the equilibrium position of an elliptical cylinder. Fig.~\ref{upper_slip_phase} shows the phase diagram for the steady motion types of the elliptical cylinder with respect to the channel-particle size ratio $\beta$ and slip length on the upper wall $L_s^{up}/d$. Different scatter points indicate different types of motion. In addition to the three types of motion that occur between no-slip walls (``lying", ``standing" and ``tumbling")~\cite{cai2024_PartA_noslip}, the unilateral slip wall leads to two new types of motion, leaning against the wall (simple called ``leaning")  and rolling on the wall (simple called ``rolling"). The first one is with the ellipse moving forward with a certain tilt angle and leaning against the wall, and the second one is with the ellipse rolling forward along the wall, as pictorially presented on Fig.~\ref{upper_slip_phase}.  
According to the phase diagram, we also present the equilibrium positions for the center of mass of the ellipse in Fig.~\ref{upper_slip_equilibrium_position}. The scatter points are consistent with the type of motion illustrated in Fig.~\ref{upper_slip_phase}.
The alteration in slip length and channel-particle size ratio gives rise to intricate modifications to the type of motion and equilibrium position.

The equilibrium position and motion type of the ellipse are primarily influenced by wall pressure and fluid shear, as evidenced by the results under no-slip boundary condition~\cite{cai2024_PartA_noslip}. The velocity distribution of Poiseuille flow with slip on the upper wall, shown in Fig.~\ref{upper_slip_poiseuille_vel}, indicates that increasing the slip length on the upper wall decreases the shear rate near the upper wall and meanwhile increasing it near the lower wall. Additionally, slip reduces the wall pressure on the particles near the upper wall by weakening the velocity gradient between the upper and lower sides of the fluid flowing around the particles.

According to the phase diagram in Fig.~\ref{upper_slip_phase} and the equilibrium position in Fig.~\ref{upper_slip_equilibrium_position}, we can categorize the results based on three distinct strengths of channel confinement and further analyze the effects of slip length on the dynamics of the ellipse. Region I represents a strong confinement with $\beta \leq 1.2$, wherein the wall force, inclusive of pressure from hydrodynamic interactions in addition to particle-wall interaction, exhibits a dominant influence with varying slip length. All five motion types occur at both $Re=10$ and $Re = 1$. The leaning and rolling types are unique to this region. The particle's equilibrium position initially increases and then decreases as slip length increases. The initial rise is largely attributed to the weakening of the upper wall force, while the subsequent decline is due to the alteration of the flow field structure, which gives rise to the shear lift of particles directed away from the upper wall surface. Region II represents moderately confined channels with $\beta = 1.4$ at $Re=10$ or $1.4 \leq \beta \leq 1.8$ at $Re=1$, where particles transition from standing motion to tumbling as slip length increases. Region III represents weakly confinement with $\beta \geq 1.6$ at $Re=10$ or $\beta \geq 2.0$ at $Re=1$, where changes in shear gradient lift dominate. Ellipses in this region exhibit tumbling motion with two equilibrium positions at small slip lengths. When the slip length exceeds a threshold, the equilibrium position is only the one on the lower side because the ellipse near the distance from the upper wall does not have enough shear gradient lift to maintain equilibrium. As the slip length continues to increase, this equilibrium position rises.

\begin{figure*}
\centering
\includegraphics[scale=0.62]{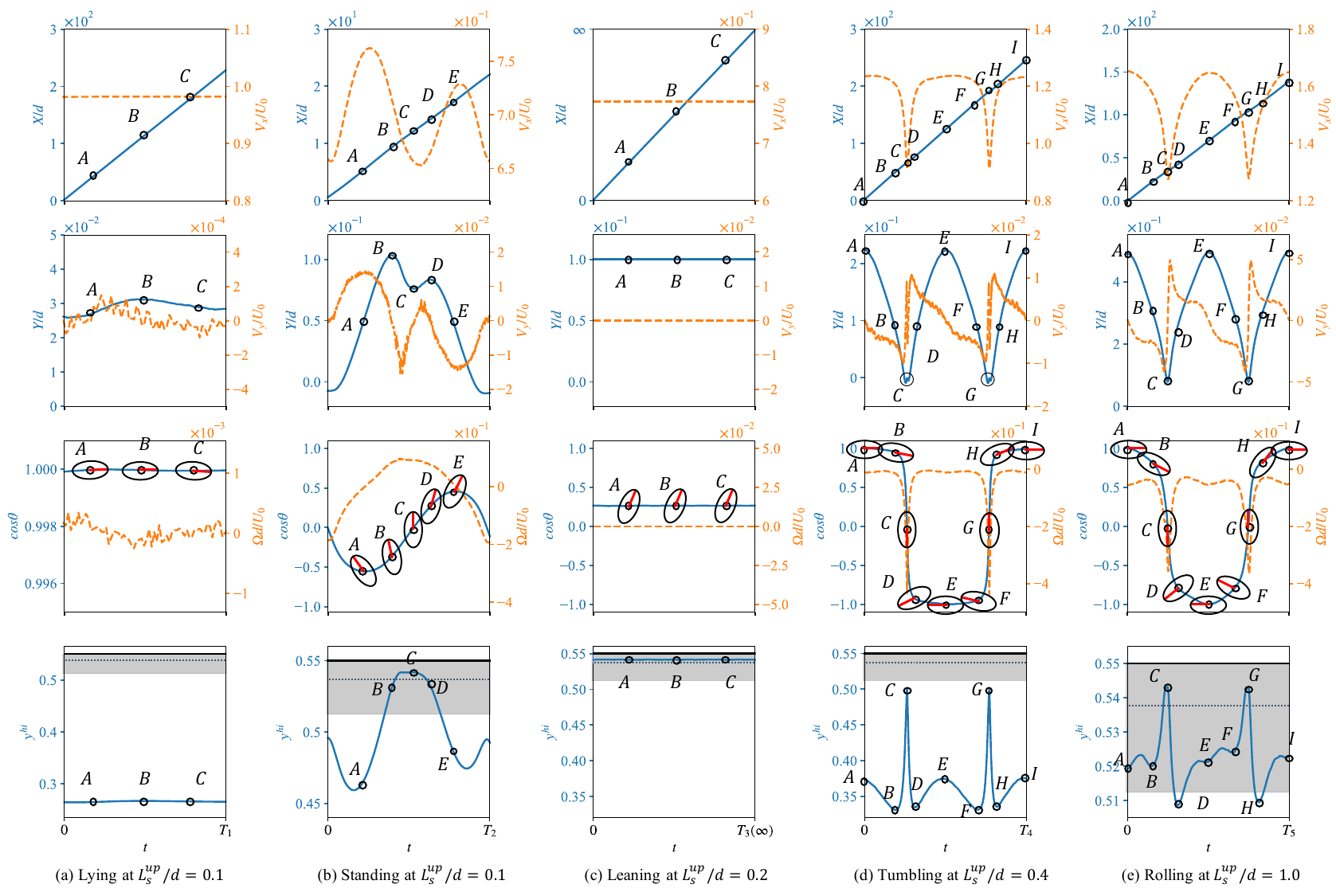}
\caption {Different types of periodic behaviors of the ellipse with $\beta=1.1$ and $Re=10$ under different upper slip lengths at steady state. Horizontal coordinate represents one full period, which has different values ($T_1-T_5$) for the five cases. (a) The lying motion at $L_s^{up}/d = 0.1$: $T_1=12.7$,
(b) The standing motion at $L_s^{up}/d = 0.1$: $T_2=1.68$,
(c) The Leaning motion at $L_s^{up}/d = 0.2$: $T_3=\infty$,
(d) The tumbling motion at  $L_s^{up}/d = 0.4$: $T_4=11.27$,
(e) The rolling motion at $L_s^{up}/d = 1.0$: $T_4=4.95$.
The gray area indicates that there is an interaction between the ellipse and the wall, with the dashed line signifying the activation of the short-range repulsive force.}
\label{upper_slip_period_motion}
\end{figure*}

Fig.~\ref{upper_slip_period_motion} illustrates the steady-state dynamics behavior of the five motion types for an ellipse at various slip lengths with channel-particle size ratio $\beta=1.1$ in one period. As slip length increases, the ellipse exhibits five motion types: lying and standing at $L_s^{up}/d=0.1$, leaning at $L_s^{up}/d=0.2$, tumbling at $L_s^{up}/d=0.1$, as well as rolling at $L_s^{up}/d=1.0$. The figure from top to bottom shows the longitudinal displacement ($X/d$) and velocity ($V_x/U_0$), lateral displacement ($Y/d$) and velocity ($V_y/U_0$), the cosine of the direction angle along $x$-axis ($\theta$) and angular velocity ($\Omega d/U_0$), as well as the minimum distance between the ellipse and the wall ($y^{in}$). The gray area indicates that particle-wall interaction is enabled in this region.
Lying and standing behaviors are similar to no slip. Leaning is a new type of motion in which the ellipse leans against the wall at a tilt angle $\theta=72^{\circ}$. In this position, the ellipse is no longer undergoing rotation, as the clockwise torque generated by the wall force is balanced by the counterclockwise torque resulting from the shear in the flow field. In the lateral direction the ellipse is balanced by the wall force and the shear gradient lift, so there is no displacement. In the longitudinal direction it moves forward with a constant velocity.
Rolling is also a new type of motion similar to the tumbling. Rolling type occurs at larger slip lengths and is subject to strong interactions with the wall due to the strongly confined environment. Throughout the majority of the cycle, the ellipse is very close to the wall and the particle-wall interaction is activated.

\subsection{Identical slip length on both walls}
\begin{figure*}
	\centering  
	\subfigure[$Re=10$]{
		\label{}
		\includegraphics[scale=0.45]{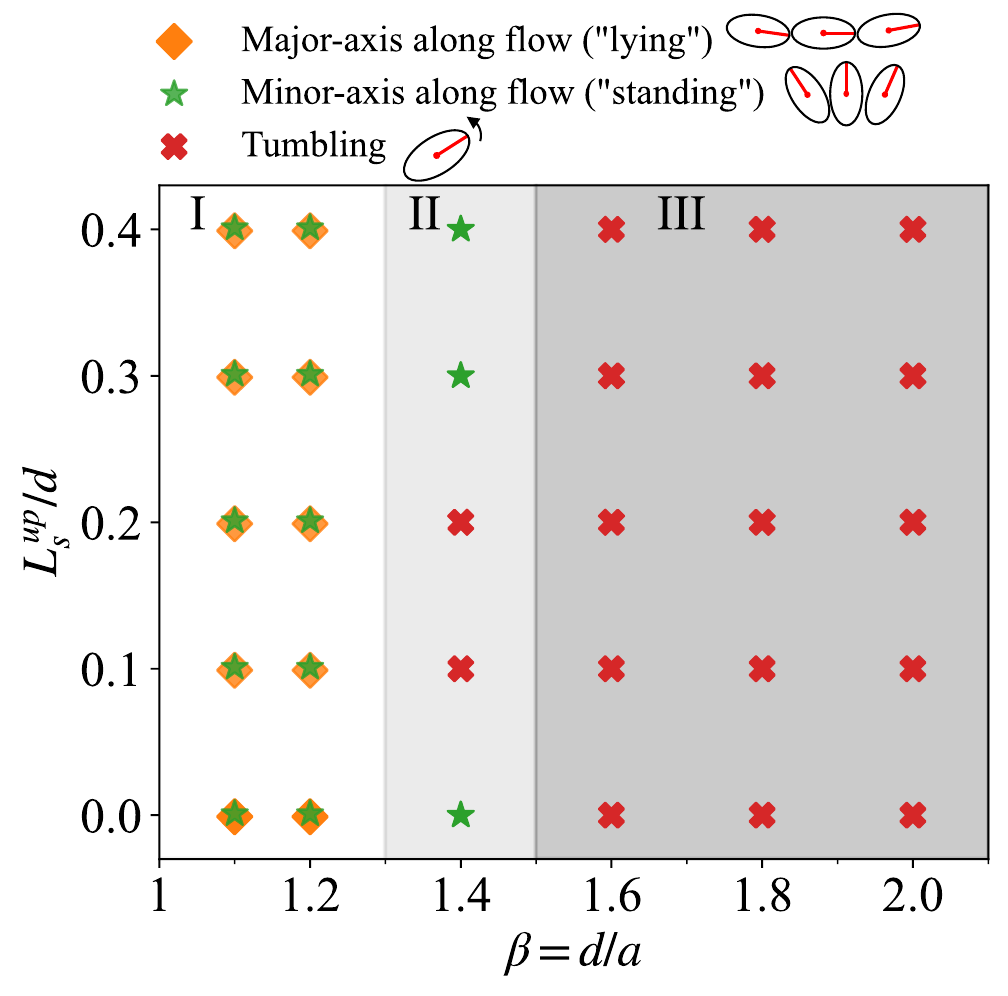}}
	\subfigure[$Re=1$]{
		\label{}
		\includegraphics[scale=0.45]{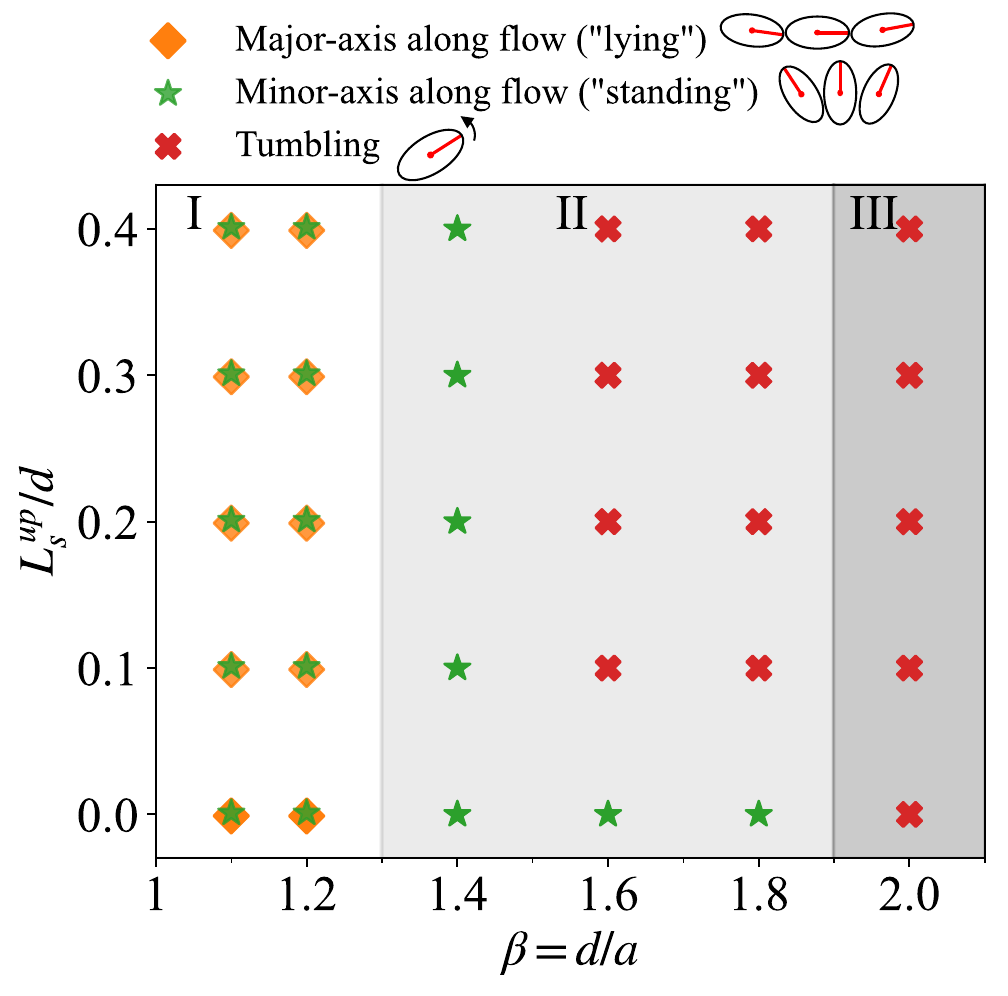}}
	\caption{Phase diagram for the steady behaviors of the ellipse with
respect to the channel–particle size ratio $\beta$ and slip length on the both walls $L_s/d$.}
	\label{both_slip_phase}
\end{figure*}

\begin{figure*}
	\centering  
	\subfigure[$Re=10$]{
		\label{}
		\includegraphics[scale=0.55]{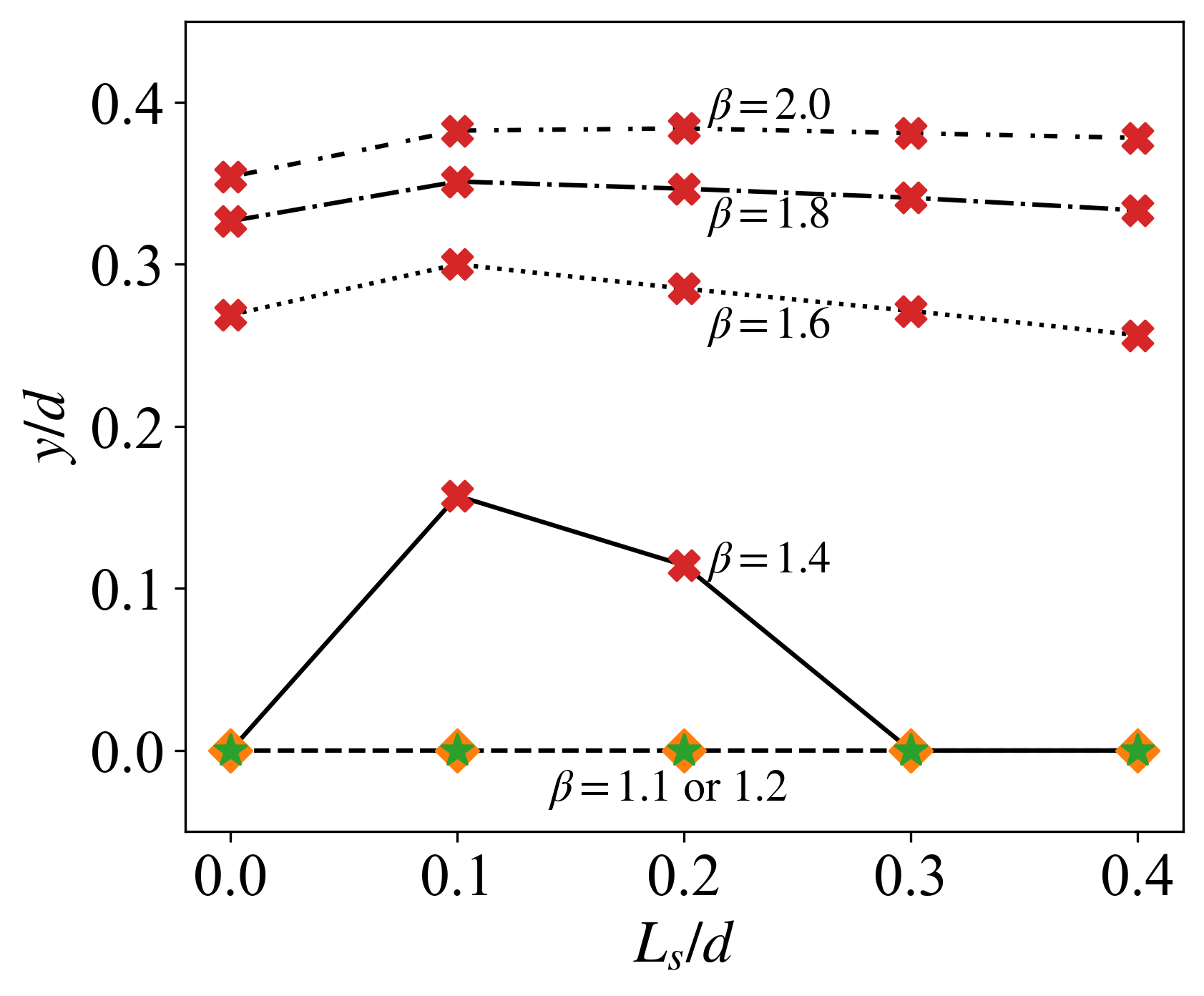}}
	\subfigure[$Re=1$]{
		\label{}
		\includegraphics[scale=0.55]{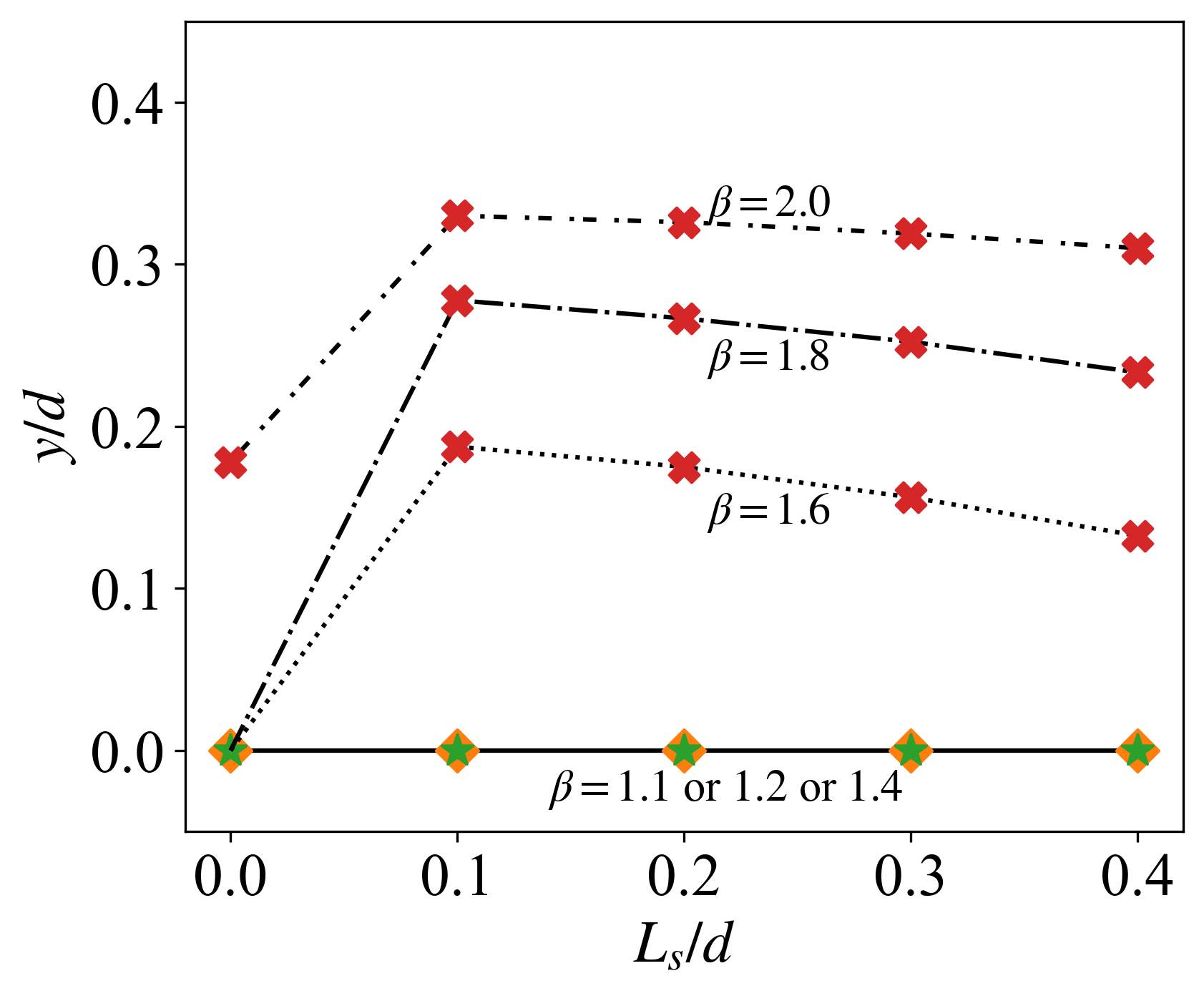}}
	\caption{Equilibrium position of the center of mass of the ellipse for different slip lengths on the both walls $L_s/d$. The scatter symbols indicate the type of motion and have the same denotation as illustrated in Fig~\ref{both_slip_phase}. }
	\label{both_slip_position}
\end{figure*}

We now examine the scenario where identical slip conditions occur at both the top and bottom walls of the channel in Poiseuille flow. Fig.~\ref{both_slip_phase} illustrates the phase diagram depicting the steady behaviors of the ellipse as a function of the channel-particle size ratio ($\beta$) and the slip length on both walls ($L_s/d$).
Two Reynolds numbers $Re = 10$ and $1$ are considered, along with five slip lengths: $ L_s/d = 0$, $0.1$, $0.2$, $0.3$, and $0.4$. 
Three types of motion are observed: lying, standing, and tumbling.  Fig.~\ref{both_slip_position} shows the equilibrium positions of particles at various slip lengths. When the focusing positions are on the centerline, the ellipse exhibits a standing or lying motion type. Conversely, when the focusing positions are off the centerline, the ellipse displays a tumbling motion type. These observations align with the no-slip results.

\begin{figure*}
	\centering  
	\subfigure[Lateral position.]{
		\label{}
		\includegraphics[scale=0.4]{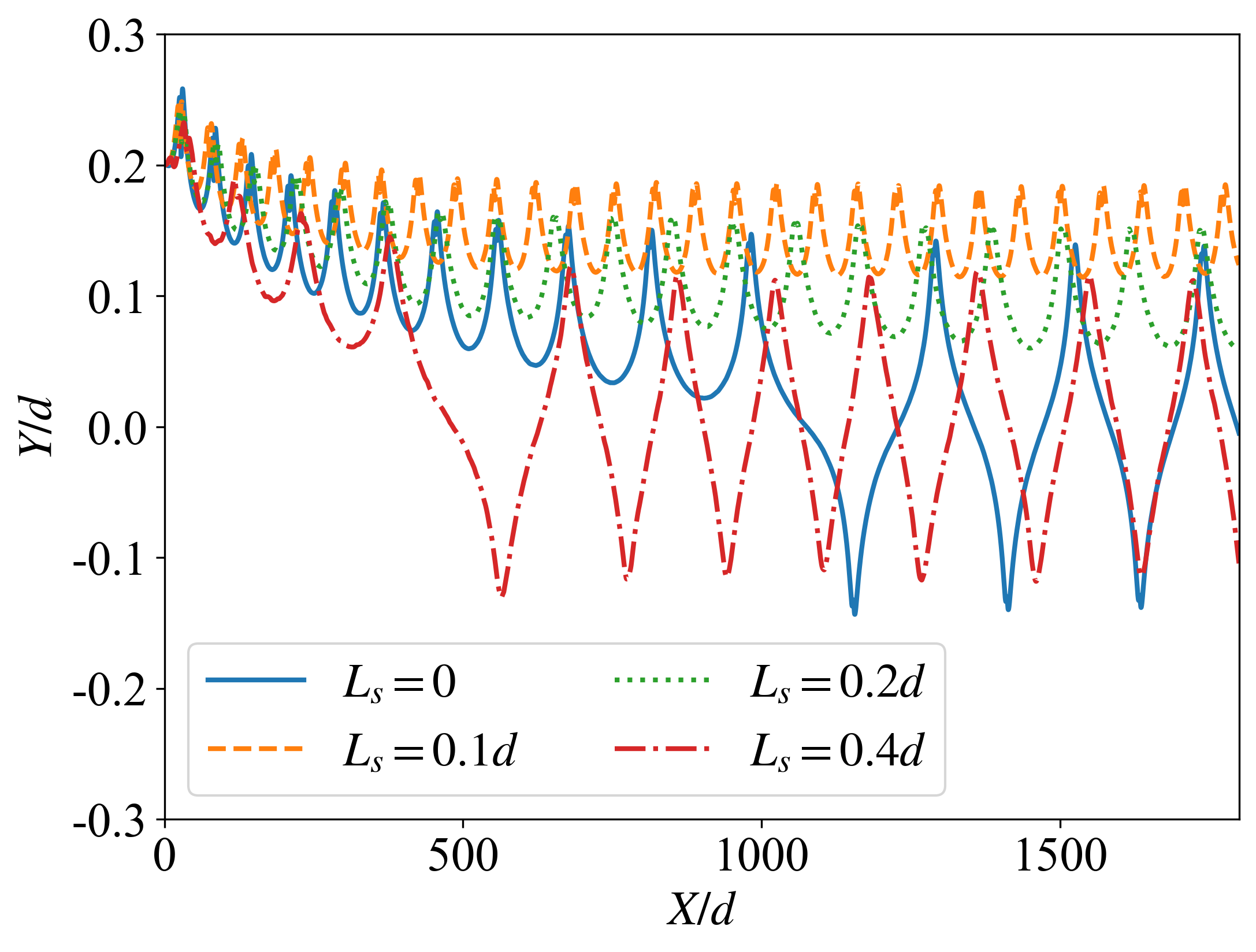}}
	\subfigure[Angle of the rotation.]{
		\label{}
		\includegraphics[scale=0.4]{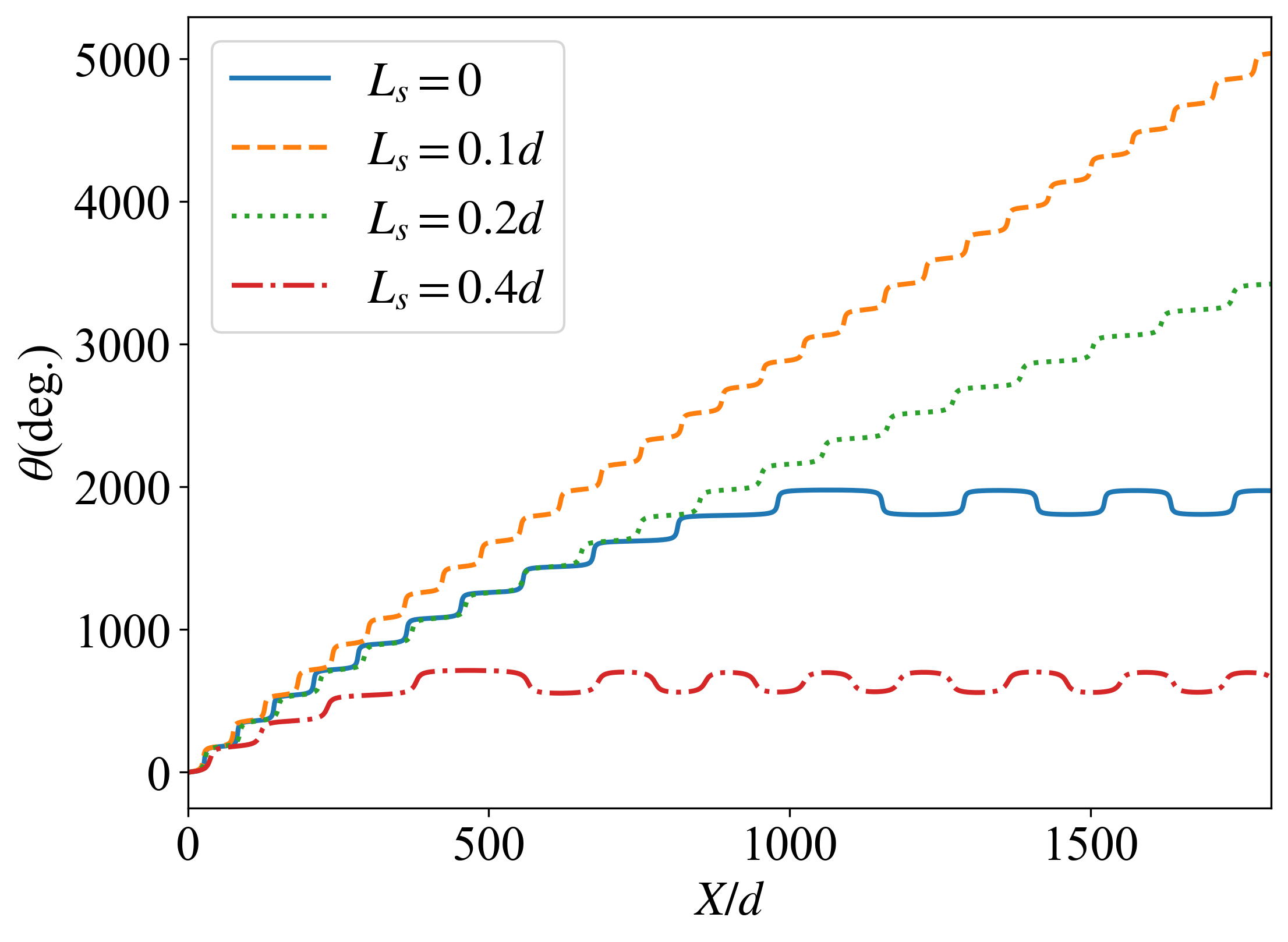}}
	\caption{The trajectory and rotation of the ellipse from rest at $Re=10$ and $\beta=1.4$.  
$X/d$ is the longitudinal displacement of the ellipse. $Y/d$ and $\theta$ are the lateral position and angle of the rotation, respectively. The initial positions of the ellipses are all located at $(Y_0,\theta_0)=(0.2d, 0)$.}
	\label{Transient_behavior_re10_beta1_4}
\end{figure*}

In Region I of the phase diagram presented in Fig.~\ref{both_slip_phase}, which corresponds to a strongly confined environment, the motion type of the ellipse is either lying or standing at steady state, depending on the initial position and/or orientation, regardless of the slip length. In Region II, particles are moderately confined by the channel, and their motion type can be either standing or tumbling, contingent upon the slip length $L_s$, the particle-channel size ratio $\beta$ and Reynolds number $Re$. In Region III, particles are in weakly confinement, resulting in tumbling motion type exclusively. 
Regarding equilibrium positions of the ellipse in Fig.~\ref{both_slip_position}, Region I corresponds to the centerline. In contrast, Regions II and III exhibit more complexity, with the equilibrium position initially approaching the wall and then the centerline as the slip length increases. This non-monotonic variation is most pronounced at $\beta=1.4$ with $Re=10$. Fig.~\ref{Transient_behavior_re10_beta1_4} shows the motion of the ellipse from rest at the same initial position $ (Y_0, \theta_0) = (0.2d, 0)$ for different slip lengths. At steady state, as the slip length increases, the particle transitions from standing motion at centerline with no slip to tumbling motion at $L_s/d = 0.1$, where the inertial focusing position is $y/d = 0.157$. At $L_s/d = 0.2$, tumbling continues, and the equilibrium position decreases to $y/d = 0.11$. After $L_s/d=0.3$, the particles return to a standing motion at the centerline. 

\begin{figure*}
	\centering  
	\subfigure[Distribution of pressure on the whole flow filed]{
		\label{beta_1_4_wall_slip_pressure_a}
		\includegraphics[scale=0.32]{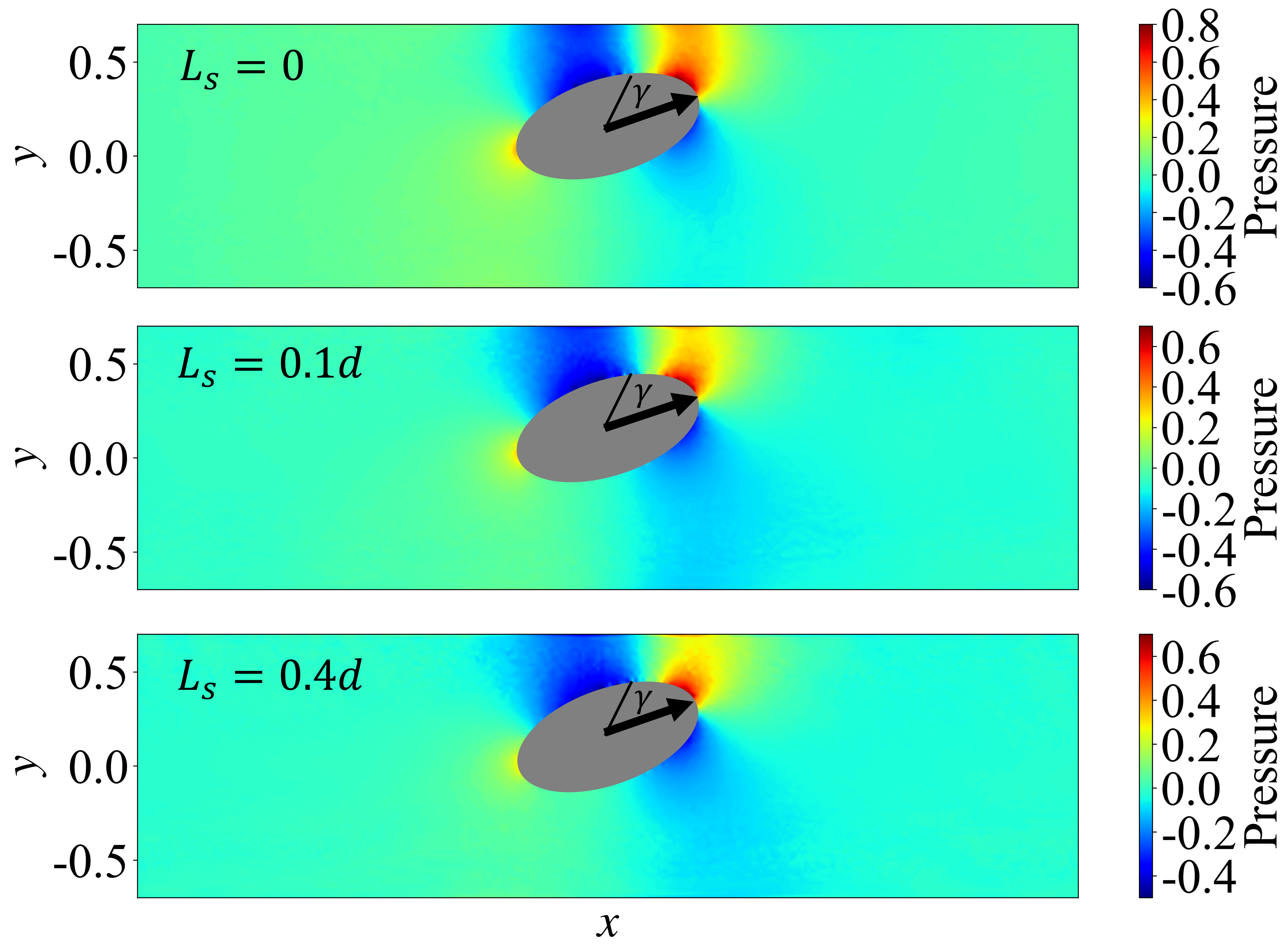}}
	\subfigure[Distribution of pressure on the surface of the ellipse]{
		\label{beta_1_4_wall_slip_pressure_b}
		\includegraphics[scale=0.42]{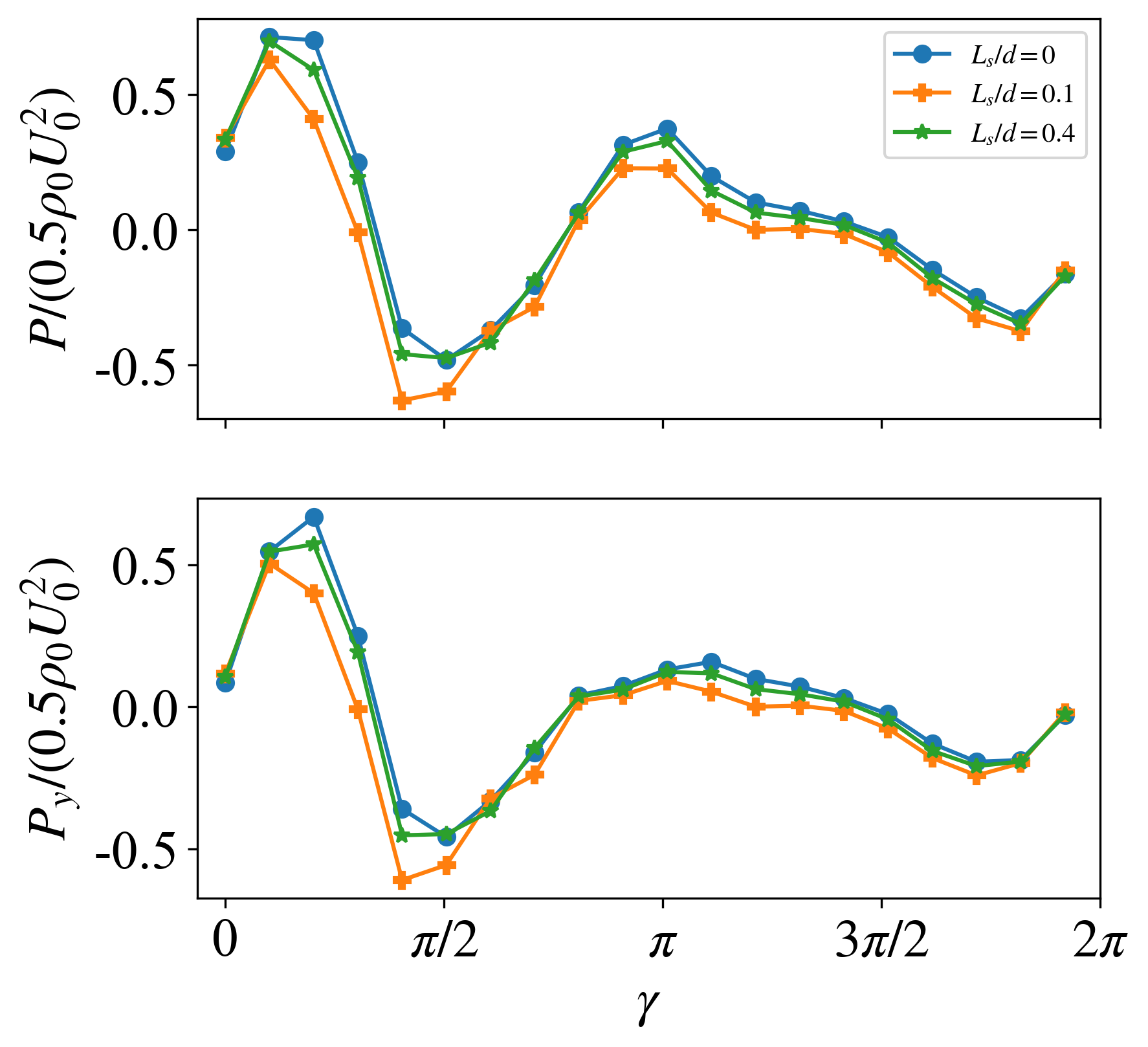}}
	\caption{Distribution of pressure on the whole flow field and on the surface of the ellipse with $\beta = 1.4$ and $Re=10$ for various slip lengths on the both walls.}
	\label{beta_1_4_wall_slip_pressure}
\end{figure*}

Variations in slip length do not significantly affect shear, but only augment the fluid velocity, as shown in the velocity distribution in Fig.~\ref{both_wall_slip_poiseuille_vel}. Consequently, the phenomena occurring in Regions II and III may be closely related to variations in wall-induced interaction force. As the slip length increases, wall-induced force initially decreases and then increases. We plotted the pressure distributions of the flow field at different slip lengths ($L_s/d = 0, 0.1, 0.4$) and the corresponding pressure distributions on the surface of the ellipse with $\beta=1.4$ and $Re=10$, as illustrated in Fig.~\ref{beta_1_4_wall_slip_pressure}.
To better analyze the pressure among these cases, the
transverse positions of the center of mass of the ellipse $Y$ and the orientation $\theta$ are closely matched. The pressures shown in Fig.~\ref{beta_1_4_wall_slip_pressure_a} are obtained via the EOS and then normalized to $P/0.5\rho_0U_0^2$. They are then interpolated to uniform meshes using a quintic kernel function. Fig.~\ref{beta_1_4_wall_slip_pressure_b} shows the pressure distribution on the surface of the ellipse along the anticlockwise direction, where the upper graph presents the pressure magnitude and the lower represents the projection of pressure in the $y$-direction. The horizontal axis $\gamma$ represents the angle between a line from points on the ellipse's surface to its centroid and the long axis (depicted by the black arrow in the Fig.~\ref{beta_1_4_wall_slip_pressure_a}). The maximum pressure occurs at the front side of the ellipse near the wall, while the smallest negative pressure is located at the rear side near the wall. A smaller slip length results in higher lateral wall pressure.

To summarize this part, under the condition of equal slip lengths ($L_s/d \leq 0.4$) on both the upper and lower walls in Poiseuille flow, a small slip length enables particles to achieve an inertial focusing position away from the centerline or to enhance their equilibrium positions. However, further increasing the slip length diminishes this enhancement. This variation with slip length is observed only in moderate and weakly channel confinement and becomes less pronounced as channel confinement weakens (or $\beta$ increases) and the Reynolds number decreases. In the case of strong channel confinement, the type of motion and equilibrium position of the particles are not influenced by the slip length.

\subsection{Slip boundary on the elliptical cylinder surface}
\begin{figure*}
	\centering  
	\subfigure[$Re=10$]{
		\label{}
		\includegraphics[scale=0.45]{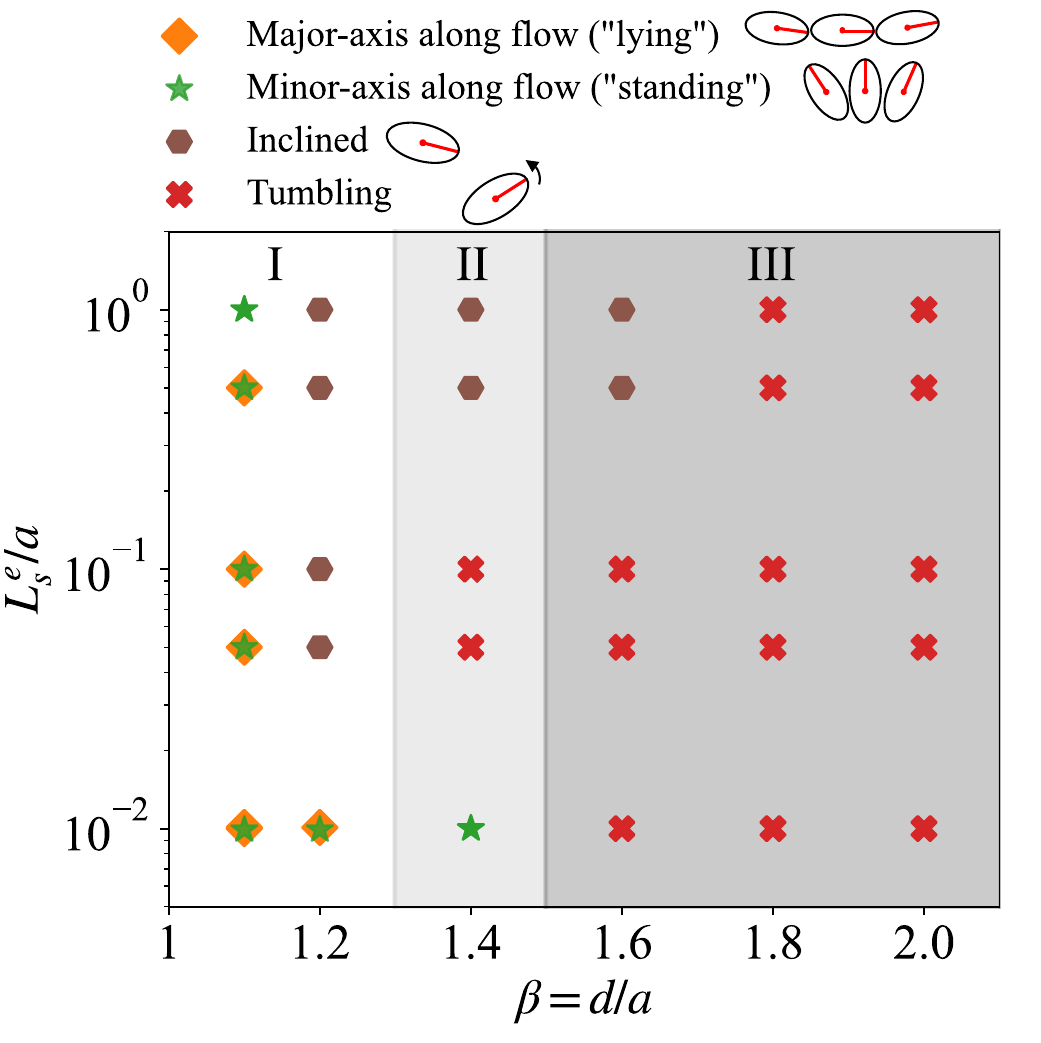}}
	\subfigure[$Re=1$]{
		\label{}
		\includegraphics[scale=0.45]{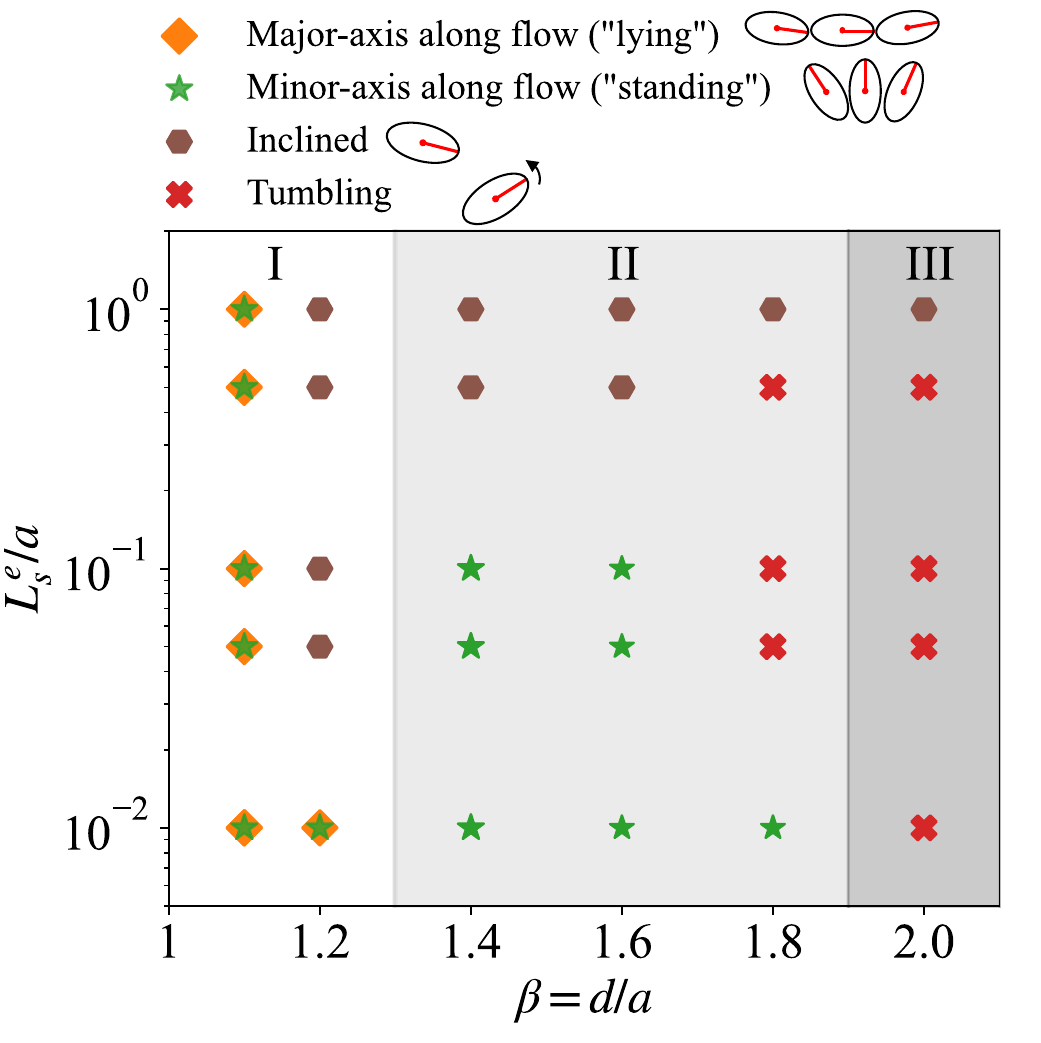}}
	\caption{Phase diagram for the steady behaviors of the ellipse with
respect to the channel–particle size ratio $\beta$ and slip length on the ellipse $L_s^e/a$.}
	\label{phase_slip_on_ellipse}
\end{figure*}

\begin{figure*}
	\centering  
	\subfigure[$Re=10$]{
		\label{}
		\includegraphics[scale=0.55]{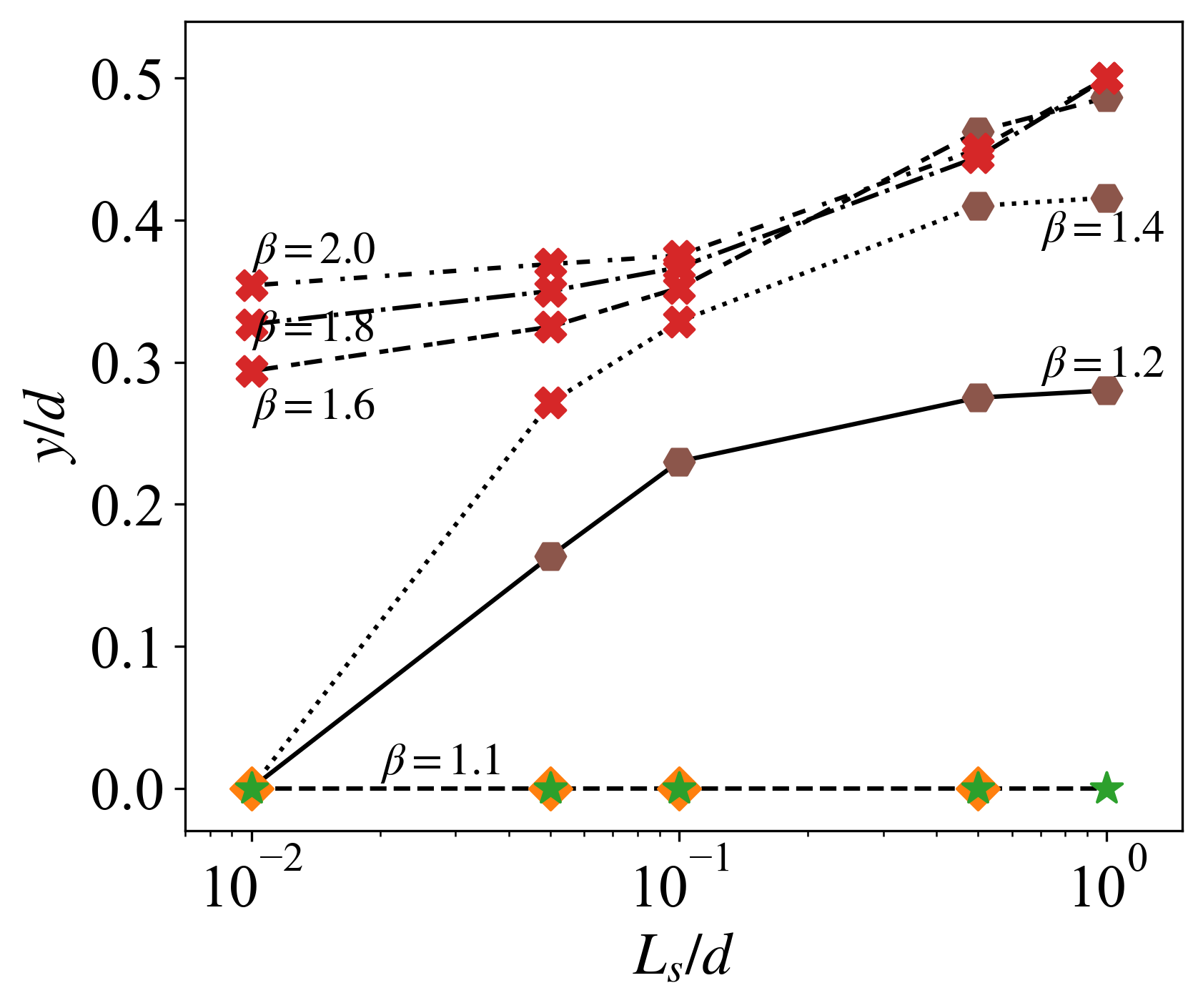}}
	\subfigure[$Re=1$]{
		\label{}
		\includegraphics[scale=0.55]{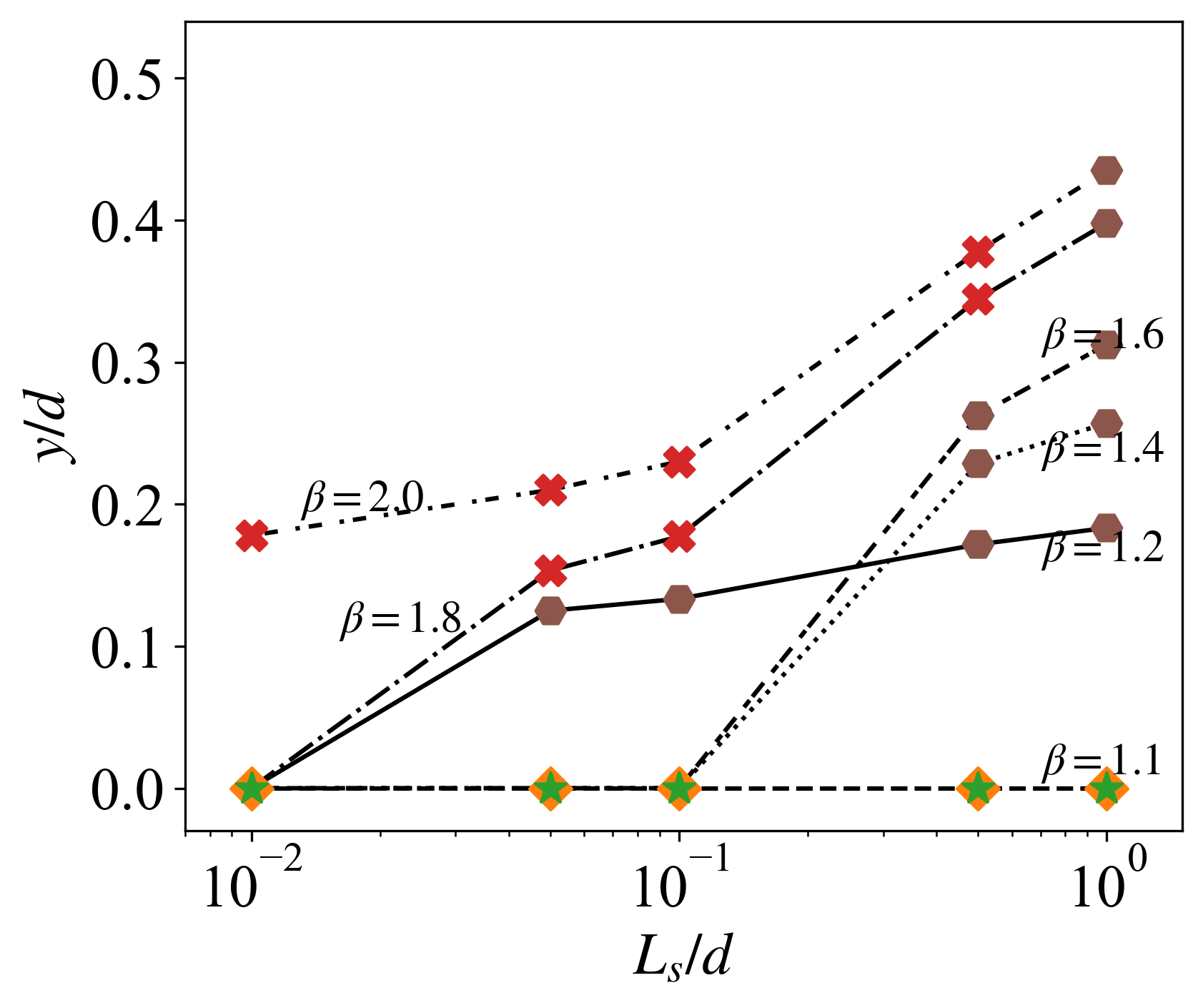}}
	\caption{Equilibrium position of the center of mass of the ellipse for different slip lengths on the ellipse $L_s^e/a$. The scatter symbols indicate the type of motion and have the same denotation as illustrated in Fig~\ref{phase_slip_on_ellipse}. }
	\label{ellipse_slip_position}
\end{figure*}

\begin{figure*}
	\centering  
	\subfigure[Lateral position.]{
		\label{}
		\includegraphics[scale=0.4]{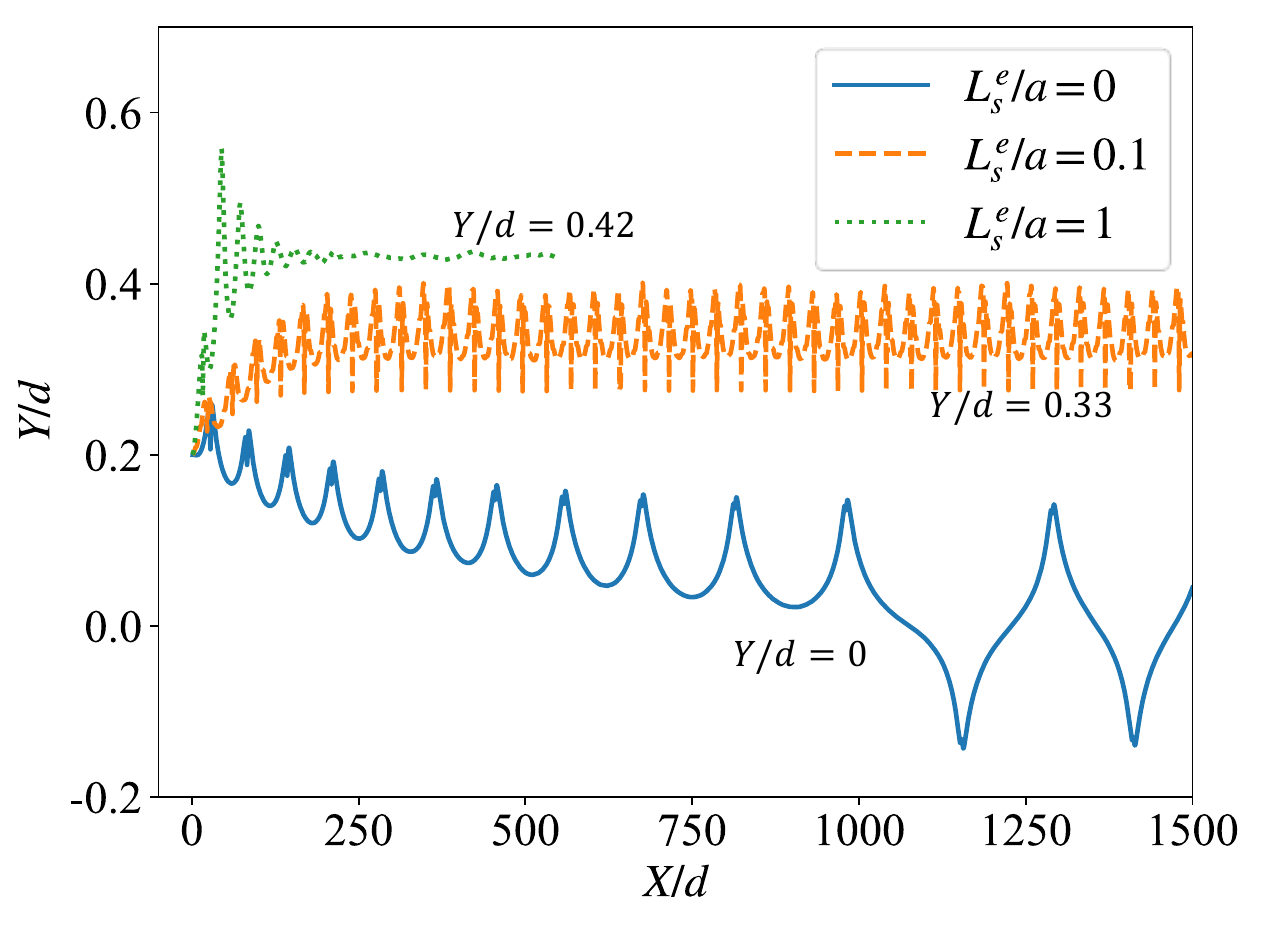}}
	\subfigure[Angle of the rotation.]{
		\label{}
		\includegraphics[scale=0.4]{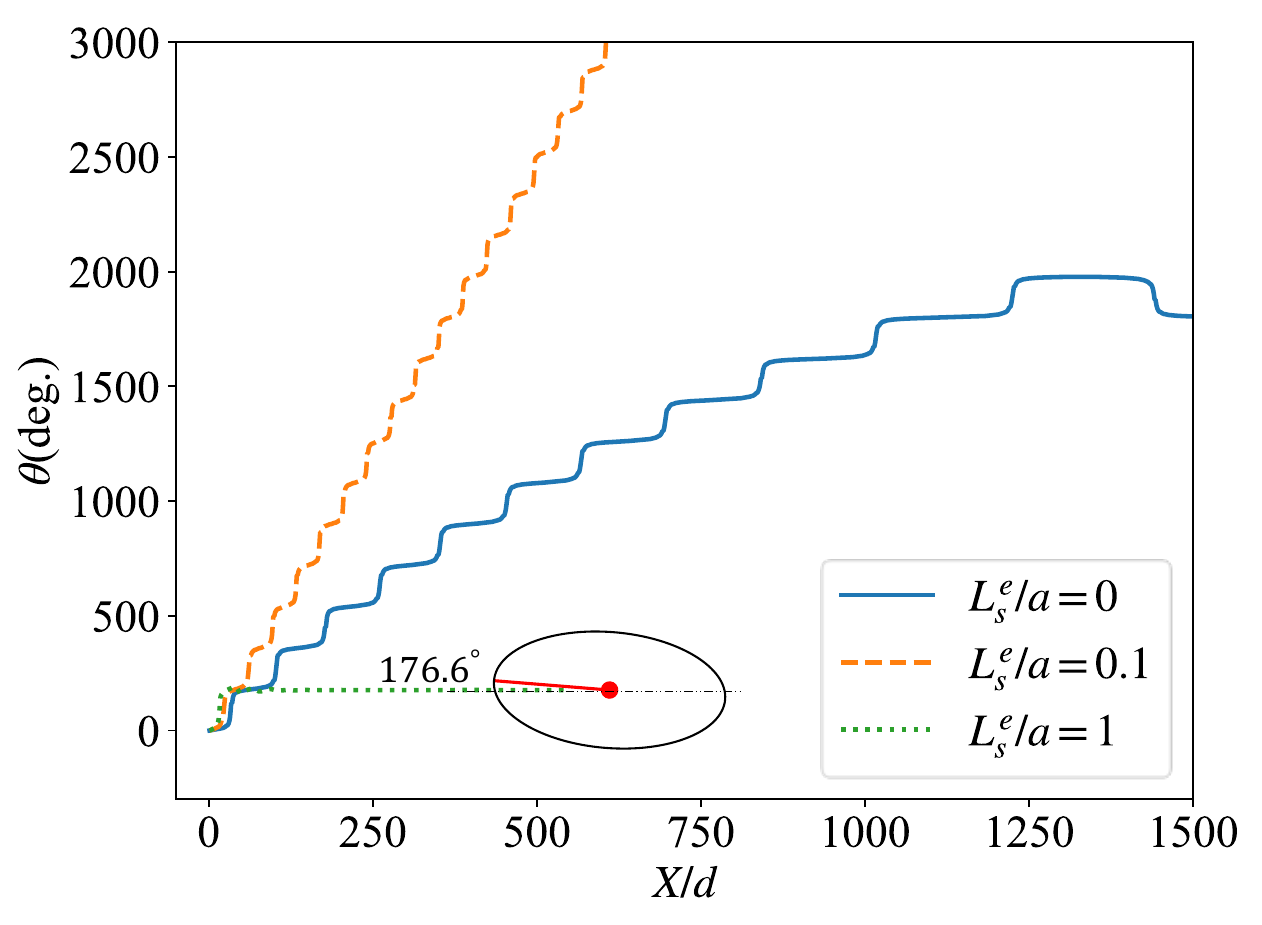}}
	\caption{The trajectory and rotation of the ellipse with different slip length on the ellipse.}
	\label{compare_slip_on_ellipse_beta1_4_re10}
\end{figure*}

Furthermore, we examine the scenario of slip occurring on the surface of the ellipse while the wall exhibits no slip. The slip length on the ellipse, denoted as $L_s^e$, is set at values of $0.01$, $0.05$, $0.1$, $0.5$, and $1$.

Fig.~\ref{phase_slip_on_ellipse} illustrates the phase diagram of the ellipse concerning slip length and size ratio at steady state, while Fig.~\ref{ellipse_slip_position} presents the corresponding equilibrium position of the ellipse. The type of motion observed with a slip length of $L_s^e/a=0.01$ aligns with the no-slip results, which we omit for brevity. In scenarios with greater confinement of the channel or larger slip lengths, a new motion type may emerge, characterized by a fixed small negative angle between the long axis of the ellipse and the positive x-axis. This inclined configuration positions the ellipse's center between the channel's centerline and the wall, known as the "inclined" type. In our simulations, the inclination angle ranges from $-7$ to $-2$ degrees. 
In contrast to the leaning motion described in the first subsection, the inclined motion type occurs with the ellipse situated away from the wall, thereby indicating the absence of particle-wall interaction.

In region I of strong confinement/small $\beta$, the lying motion type ceases to exist when $\beta = 1.1$ and the slip length is $L_s/a = 1$. By contrast, at $\beta = 1.2$, the inclined type manifests when the slip is at least $0.5a$, with the equilibrium position approaching the wall as slip length increases. In region II, characterized by moderate channel confinement, tumbling motion is observed with increasing slip, as seen in cases with $\beta = 1.4, Re = 10$ and $\beta = 1.8, Re = 1$. The inclined motion type emerges for larger slip lengths. Fig.~\ref{compare_slip_on_ellipse_beta1_4_re10} depicts the trajectory and rotation angle of the ellipse, starting from rest at $\beta = 1.4, Re = 10$. We compare the tumbling type at steady state with  $L_s^e/a = 0.1$ and the inclined type with $L_s^e/a = 1$, against the no-slip condition. The equilibrium position of the tumbling type is $Y/d = 0.33$. For the inclined type, the ellipse rotates counterclockwise for $176.6 $ degrees before stabilizing, with oscillations in the transverse direction ceasing. The final equilibrium position is $Y/d = 0.42$. Under weakly confinement/larger $\beta$ of region III, all motion types are tumbling when the slip length is small, with steady-state equilibrium positions approaching the wall as slip length increases. As the slip length continues to rise, inclined motion types arise, such as at $\beta = 1.6, Re = 10$ and $\beta = 2.0, Re = 1$. No inclined motion type is present when $\beta \geq 1.8, Re = 10$.

The slip length on the elliptical solid reduces the tangential hydrodynamic stress on the particle surface, resulting in lowered translational drag coefficients~\cite{kamal2021effect}. As the slip length increases, the velocity lag between the particle and surrounding fluid also increases, leading to positive lag velocities at sufficiently high slip lengths~\cite{cai2023_li_xin_JCompPhys}. Consequently, this enhances the particle's translational velocity and leads to increased inertial lift. Three scenarios may arise: the first being the disappearance of lying motion (e.g., $\beta = 1.1, Re = 10$); the second involves a transition from a centered position to tumbling away from the centerline (e.g., $\beta = 1.4, Re = 10$); and the third is the increase of the equilibrium position when the particle is tumbling. (e.g., $\beta = 1.8, Re = 10$).

\begin{figure*}
\centering
\includegraphics[scale=0.35]{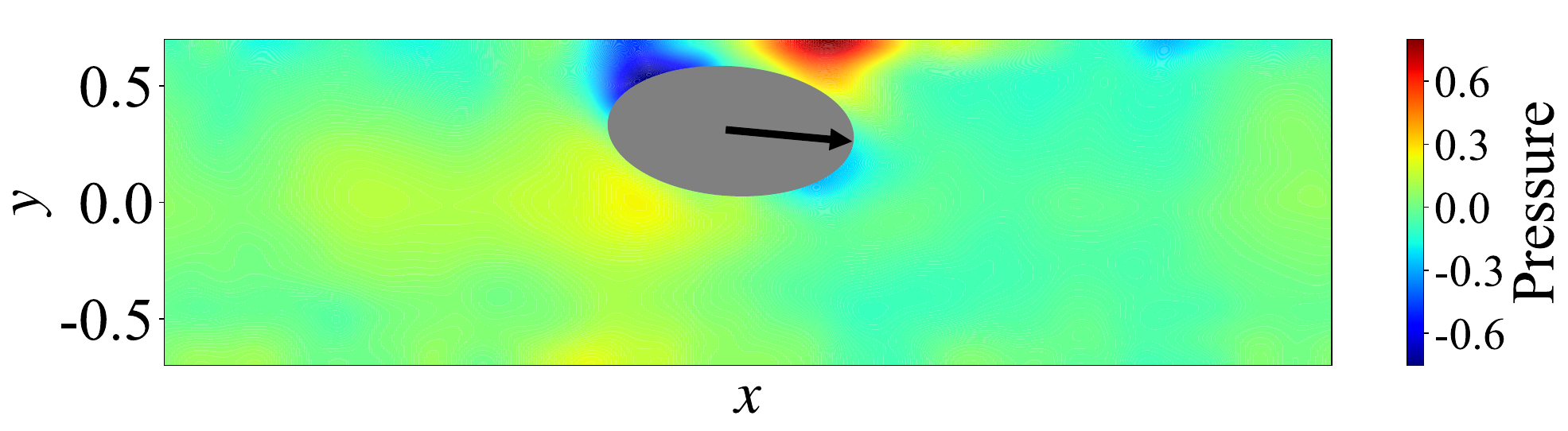}
\caption {Pressure distribution for an ellipse in Poiseuille flow with a surface slip length $L_s^e = a$, $\beta = 1.4$ and $Re=10$.}
\label{beta_1_4_ellipse_slip_pressure_re10}
\end{figure*}

Moreover, slip can also lead to a weakening of the rotational dynamics of freely suspended particles in shear~\cite{Luo2008_Pozrikidis_}. Fig.~\ref{beta_1_4_ellipse_slip_pressure_re10} shows the steady state flow field pressure distribution for an ellipse with a surface slip length $L_s^e/a = 1$ in a channel with a size ratio of $\beta=1.4$ and $Re=10$. 
The ellipse experiences two torques at this location: a clockwise torque caused by the wall effect, resulting from positive pressure on the front side of the ellipse near the wall and negative pressure on the back side, and a counterclockwise torque due to the shear. At this specific angle, the two torque reach equilibrium.
This phenomenon is akin to the findings reported by Kamal et al. (2020)~\cite{kamal2020hydrodynamic}, where they conducted a study in which rigid graphene-like nanoparticles in shear flow with large slip boundary conditions were modelled using a combination of molecular dynamics and boundary integral simulations and found that the platelet was aligned at small angles relative to the flow direction instead of exhibiting a periodically rotating Jeffery orbit.
Li and Ardekani~\cite{Li2014_Ardekani_PhysRevE} showed that a spherical squirmer with a tangential surface velocity develops a similar phenomenon of inclined mode near the wall, traveling along the wall at a constant distance with a negative angle.

\section{Conclusion}\label{conclusion}

In this study, we employ the smoothed particle hydrodynamics method, which incorporates a novel approach for boundary treatment, to simulate the dynamics of an elliptical cylinder in confined Poiseuille flow with slip boundary conditions. By modifying the wall slippage or the ellipse surface wettability, we illustrate the capacity to influence the inertial focusing position or the motion modes of the particle, thereby actively controlling the particle dynamics.

Adjusting the slip length on the upper wall alters the velocity distribution, thereby changes the velocity gradient and wall-induced pressure.
Compared to the no-slip scenarios, two novel motion modes termed leaning and rolling are observed when slip is applied to the unilateral wall. 

When equal slip lengths are applied to both walls in the moderate and weakly confinement ($\beta \leq 1.4$), a small slip length allows the particle to achieve an inertial focusing position away from the centerline and elevate the equilibrium position. However, further increasing the slip length diminishes this enhancement.

The application of slip length on the elliptical surface decreases the tangential hydrodynamic stress and impacts the rotational dynamics. The reduction in tangential stress increases particle inertial lift, while the diminished shear torque modulates the particle's motion, leading to the new motion type termed as inclined motion.

\section*{Acknowledgments}
The authors acknowledge the national natural science foundation of China under grant number: 12172330,
and the grant of the national key R\&D program of China under contract number 2022YFA1203200.


\bibliographystyle{unsrt} 
\bibliography{main} 

\end{multicols}


\makeentitle

\end{document}